    \documentclass[
        universe,
        article,
        accept,
        pdftex,
        moreauthors
    ]{Definitions/mdpi}

    \firstpage{1}
    \pubvolume{1}
    \issuenum{1}
    \articlenumber{0}
    \pubyear{2026}
    \copyrightyear{2026}
    \datereceived{ }
    \daterevised{ } 
    \dateaccepted{ }
    \datepublished{ }

    \usepackage{aas_macros}
    \newcommand*{\rd}[2]{\frac{\mathrm{d}#1}{\mathrm{d}#2}}
    
    \newcommand*{\rdil}[2]{\mathrm{d}#1 / \mathrm{d}#2}
    
    \newcommand*{\bm}[1]{\boldsymbol{\mathbf{#1}}}
    \newcommand*{\uv}[1]{\hat{\bm{#1}}}
    \newcommand*{\at}[1]{\left.#1\right|}
    \newcommand*{\abs}[1]{\left|#1\right|}
    \newcommand*{\ev}[1]{\left\langle#1\right\rangle}
    \newcommand*{\p}[1]{\left(#1\right)}
    \newcommand*{\s}[1]{\left[#1\right]}
    \newcommand*{\z}[1]{\left\{#1\right\}}
    \newcommand{\jhat}{\uv{j}}

\Title{
A Possible Path to Mass Ratio and Spin-Orbit Misalignment Correlation: Mergers
of Binary Black Holes in Nuclear Star Clusters
}

\Author{Yubo Su $^{1,2,*}$\orcidA{}}

\AuthorNames{Yubo Su}

\address{%
$^{1}$ \quad Canadian Institute for Theoretical Astrophysics, 60 St.\ George
Street, Toronto, ON M5S 3H8, Canada\\
$^{2}$ \quad Department of Astrophysical Sciences, Princeton University,
Princeton, NJ 08544, USA
}

\corres{Correspondence: yubo.su@utoronto.ca}

\abstract{
Despite a decade's worth of gravitational wave observations, the origin of the
binary black hole (BBH) mergers detected by the LIGO-Virgo-KAGRA (LVK)
collaboration remains an open question.
Towards assessing the feasibility and prevalence of the many proposed BBH
formation channels, the spin properties of the merging black holes (BHs) hold
significant promise, particularly their orientations.
Moderate preferential alignment of BH spins with their orbit normals, as well as
an apparent anti-correlation between the BBH effective spin parameters
$\chi_{\rm eff}$ and their mass ratios, would appear to disfavor purely
dynamical BBH formation mechanisms, as they lack a preferred orientation to the
system.
However, alternatives are filled with physical and modeling uncertainties.
In this paper, we highlight a dynamical route to easily characterizable spin
evolution that results in analytically predictable spin distributions.
We show that, when a stellar binary forms a BBH through two phases of stable
mass transfer, and the BBH is subsequently driven to merger by the gravitational
perturbation of a distant massive object (such as a supermassive black hole),
the resulting $\chi_{\rm eff}$
may be anti-correlated with the binary mass ratio.
While the mechanism as proposed only operates in a modest region of parameter
space, it also predicts significantly tighter correlations than are seen in the
GWTC-4.0 systems.
We discuss avenues for future work that may significantly expand the parameter
space of our mechanism while still remaining broadly consistent with
observations.
}


\begin{document}

\section{Introduction}\label{s:intro}

With the conclusion of the first part of the fourth LIGO-Virgo-KAGRA (LVK)
observing run, $218$ detected events have likely astrophysical origins
\citep{ligoo4a_updating}, and the large majority of these are thought to be due
to merging binary black holes (BBHs).
Many properties of these systems can be discerned from their gravitational wave
(GW) waveform, such as the binaries' component masses, mass ratios, and spin
properties.
In principle, these properties should carry information about the formation
mechanism(s) of BBHs.
In practice, such an inference has proven difficult despite extensive efforts.
Broadly, three classes of formation channels can be identified.
The first is the class of isolated BBH formation, where an isolated stellar
binary evolves into a merging BBH in isolation from any additional
perturbations \citep[e.g.,][]{lipunov1997black, podsiadlowski2003formation,
belczynski2010effect, dominik2012double, belczynski2016first, mandel2016merging,
de2016chemically, marchant2016new, belczynski2020_eff, riley2021chemically,
posydon_code, marchant2024_massivereview, posydon2_code}.
The second is the class of hydrodynamically-assisted BBH mergers, typically in
the disks of active galactic nuclei (AGN), where the BBH is formed and/or driven
towards merger due to interactions with the dense surrounding gas
\citep[e.g.,][]{mckernan2012, stone2017, leigh2018rate, secunda2019,
tagawa2020formation, li2021, li2022, li2022b, samsing2022agn, mckernan2022_qchi,
rowan2023agn, whitehead2024agn, mckernan2024_mcfacts}.
The third is the class of dynamically-assisted BBH mergers, where an initially
wide binary is induced to merge via interactions with either a few
\citep[e.g.,][]{miller2002four, wen2003eccentricity, antonini2012_nsczlk,
Antonini_2014, silsbee2017lidov, LL17, LL18,
randall2018induced, hoang2018black,
fragione2019loeb, bin_misc5, LL19, LLW_apjl, bin_misc2, su2021_lk90,
michaely2020,liu2021hierarchical, su2021_massratio, martinez2021mass,
grishin2022, su2024_superthermal}
or many \citep[e.g.,][]{zwart1999black, o2006binary, miller2009mergers,
banerjee2010stellar, downing2010compact, ziosi2014dynamics, rodriguez2015binary,
antonini2016_gn, samsing2017assembly, samsing2018black, rodriguez2018post,
gondan2018eccentric, barber2024_starclusternbody, bruel2024_greatballsfire}
bodies, including many studies on BBH mergers in nuclear star clusters
(e.g.\ \citealp{antonini2010_NSCZLK, antonini2012_nsczlk, leigh2016_mergers,
leigh2018rate, hoang2018black, fragionegrishin2019_GN, chattopadhyay2023}; see
\citealp{arcasedda2023_spins} for a review).
The question of which of these channels contributes what fraction of the
observed BBH mergers remains an open question \citep{zevin2021,
costa2023_review}, and there is an ever-expanding effort to understand
both the predicted properties of BBHs formed via each formation mechanism and
the observed properties of the merging systems in the LVK data.

One strong discriminant among these channels is the spin orientation of the
merging BBHs.
In previous iterations of the LVK data releases, the best constraints on BBH
spin properties came from the parameter \citep{ligo_o3b_props}
\begin{equation}
    \chi_{\rm eff}
        \equiv
            \frac{m_1\chi_1 \cos\theta_1 + m_2\chi_2\cos\theta_2}{m_1 + m_2},
        \label{eq:chieff}
\end{equation}
and the somewhat less intuitive $\chi_{\rm p}$ parameter
\citep{schmidt2015towards}.
Here, the $m_i$ are the BH masses, $\chi_i$ their dimensionless spin magnitudes,
and $\theta_i$ the angles between their spin axes and the binary orbit normal.
With the recent release of the fourth Gravitational Wave Transient Catalog
\citep[GWTC-4.0][]{ligoo4a, ligoo4a_props},
direct constraints on the spin properties of the component black holes (BHs) are
available.
Reassuringly, the salient features of these constraints are in broad agreement
with studies performed on the GWTC-3 catalog \citep{ligo_o3b_props}.
A few of those relevant to the current work are:
(i) BH spin orientations are broadly distributed, but with a positive bias
\citep{ligo_o3b_props, callister2024_paramfree, heinzelbisco2024_60peak}---the
GWTC-4.0 constraint is that $\sim 34\%$ of BBH systems have $\chi_{\rm eff} <
0$.
(ii) The mass ratios $q \equiv m_2 / m_1$ and $\chi_{\rm eff}$ appear to be
correlated.
However, there is some debate whether this correlation is the initially-reported
$q$-$\chi_{\rm eff}$ anticorrelation \citep{callister2021_qchi, ligo_o3b_props}
or instead a broadening of $\chi_{\rm eff}$ with decreasing $q$
\citep{heinzelbisco2024_60peak}.
GWTC-4.0 does not substantially resolve this ambiguity \citep{ligoo4a_props}.

The natural next question is: which BBH formation channels satisfy these two
constraints?
Isolated evolution is manifestly capable of producing BBHs with aligned spins,
and more recent work suggests that sufficient spin-orbit misalignment can also
be generated via isolated evolution to ensure consistency with the LVK data
\citep[e.g.,][]{steinle2021_misalignisolated, banerjee2024_olejak_qchi}.
BBH formation in AGN disks can also introduce a preferred orientation for the
spins and orbit of the BBH, leading to preferential spin-orbit alignment
\citep[e.g.,][]{wang2021_agn, cook2024_mcfacts2}.
However, studies generally find that BBHs experiencing dynamically violent
evolution result in roughly isotropic distributions of $\theta$
(e.g., \citealp{antonini2017binary, LL18, rodriguez2018post, LL19, yu2020_spin,
fragione2020, su2021_lk90, arcasedda2023_spins}, but see
\citealp{kiroglu2025_gcspin}).
As such, at first glance, it would appear that dynamically-induced BBH mergers
are disfavored by the GWTC-4.0 constraints on the spin properties of merging
BBHs.

This is not necessarily the case.
Within the class of dynamically-driven BBH merger channels, those involving just
a single tertiary companion (``tertiary-induced mergers'') can produce sharp
features in the distribution of BH spin orientations under the right
conditions---this was first pointed out in the numerical work as a ``$90^\circ$
attractor'' of spin-orbit misalignment \citep{LL18, yu2020_spin, su2021_lk90}.
Subsequent work showed that this attractor arises due to a dynamical invariant
linking the initial and final spin orientations of the BHs \citep{su2021_lk90,
stegmann2026_triples}.
Importantly, this process prefers specific spin orientations despite being a
dynamically-driven BBH merger channel.
Note that the direct connection to $\chi_{\rm eff}$ is not immediate, as
$\chi_{\rm eff}$ depends on the combination of the spin-orbit misalignment and
the BH spin magnitudes.
In this work, we will focus on predictions of $\cos \theta_i$ alone.
Several studies \citep[e.g.,][]{olejak2024_unequalmass,
banerjee2024_olejak_qchi, wang2026_spinmagqchieff} have considered
the role of spin magnitude on the $q$-$\chi_{\rm eff}$ anti-correlation.

In this work, we suggest that tertiary-induced mergers, specifically driven by a
supermassive BH (SMBH) tertiary, may be able to reproduce many of the features
of the spin distribution observed by LVK\@.
Notably, the resulting spin-orbit misalignment angles are preferentially
prograde, and their cosines may be anti-correlated with the mass ratio of the
BBH\@.
In Section~\ref{s:methods}, we review the orbital and spin evolution of a BBH in
the tertiary-induced merger channel, including the dynamical invariant that
gives rise to the spin attractor, and we discuss our semi-analytical model of
the binary stellar evolution giving rise to the BBHs we consider.
In Section~\ref{s:signatures}, we show that the combined stellar and dynamical
evolution can give rise to unexpected correlations in the spin properties of
merging BBHs, and we discuss the feasibility and efficiency of our
mechanism when taking into account the nuclear star cluster that typically
surrounds SMBHs.
We discuss and conclude in Section~\ref{s:discussion}.

\section{Methods}\label{s:methods}

\subsection{Tertiary-Induced Mergers and Spin Dynamics}\label{s:spindyn}

In a tertiary-induced merger channel, two black holes orbit each other on a
compact orbit while they together are in a distant orbit with a tertiary
companion.
The orbit of the two inner black holes, with masses $m_1$ and $m_2$, is
described by the Keplerian orbital elements $a_{\rm in}$, $e_{\rm in}$, $I_{\rm
in}$, $\omega_{\rm in}$, and $\Omega_{\rm in}$, corresponding to the binary's
semi-major axis, eccentricity, inclination, argument of pericenter, and
longitude of the ascending node respectively.
The outer orbit, of the two black holes about the tertiary with mass $m_3$, is
analogously described by $a_{\rm out}$, $e_{\rm out}$, $I_{\rm out}$,
$\omega_{\rm out}$, and $\Omega_{\rm out}$.
The reference frame is oriented with the total angular momentum pointed along
the polar axis (which is nearly equal to the outer orbit's angular momentum).
Finally, call $I$ the mutual inclination between the two orbits.

Without an outer companion, the inner binary will merge due to emission of GWs
on a characteristic timescale (assuming a circular orbit, e.g.\ \citealp{LL18})
of
\begin{align}
    T_{\rm m, 0}
        &\equiv \frac{5c^5a_{\rm in}^4}{256G^3m_{12}^2\mu_{\rm in}}\nonumber\\
        &\simeq
            10 \;\mathrm{Gyr}
                \p{\frac{m_{12}}{100 M_{\odot}}}^{-2}
                \p{\frac{\mu_{\rm in}}{25M_{\odot}}}^{-1}
                \p{\frac{a_{\rm in}}{0.3\;\mathrm{au}}}^4
                ,\label{eq:Tmerger}
\end{align}
where $m_{12} = m_1 + m_2$ and $\mu_{\rm in} = m_1m_2/m_{12}$ is the reduced mass.
Thus, binaries with $a_{\rm in} \gtrsim 0.3\;\mathrm{au}$ will take longer than
a Hubble time ($10^{10}\;\mathrm{yrs}$) to merge in isolation.
Such binaries can nevertheless be induced to merge due to gravitational
interactions with their tertiary companion---such mergers are termed
``tertiary-induced mergers''.
In this section, we review the orbital and spin evolution of the inner black
holes ($m_1$ and $m_2$) in a tertiary-induced merger.

\subsubsection{Orbital Evolution: von Zeipel-Lidov-Kozai Effect}

The tertiary companion can accelerate the merger of the inner binary by exciting
its eccentricity due to the von Zeipel-Lidov-Kozai \citep[ZLK,][]{zeipel, lidov,
kozai} effect: when the inner and outer orbits are misaligned (when the mutual
inclination $I$ is between $\sim 39^\circ$ and $\sim 141^\circ$), the eccentricity
and inclination of the inner binary will oscillate.
As $I \to 90^\circ$, the maximum of these eccentricity oscillations can approach
near-unity, resulting in efficient GW radiation at each pericenter passage
\citep[e.g.,][]{liu2015suppression, naoz2016eccentric}.

The orbital evolution of the two binaries is due to a combination of Newtonian
and General Relativistic (GR) effects.
We implement the Newtonian evolution of the inner and outer binaries by
expanding their mutual gravitational interaction to the octupole order,
double-averaging (over both the inner and outer orbits), and including the
leading-order corrections to the double-averaged approximation via ``Brown's
Hamiltonian'' \citep[as given by Eq.~64 of][]{tremaine2023brown}.
The conservative components of the equations of motion are most succinctly
derived from the interaction potential $\Phi$ between the two orbits using the
Milankovitch equations \citep[e.g.,][]{tremaine2009laplace,
liu2015suppression}; in conjunction with the dissipative effect of gravitational
wave (GW) emission, the equations of motion are:
\begin{align}
    \Phi ={}& \Phi_{\rm quad} + \Phi_{\rm oct} + \Phi_{\rm B}
        + \Phi_{\rm GR},\label{eq:phis}\\
    \rd{\bm{L}_{\rm in}}{t}
        &= -\s{\bm{j}_{\rm in} \times \nabla_{\bm{j}_{\rm in}}\Phi
            + \bm{e}_{\rm in} \times \nabla_{\bm{e}_{\rm in}}\Phi}
            + \at{\rd{\bm{L}_{\rm in}}{t}}_{\rm GW}
            ,
            \label{eq:milankovich_L}
            \\
    \rd{\bm{e}_{\rm in}}{t}
        &= -\frac{1}{L_{\rm in}}
            \s{\bm{j}_{\rm in} \times \nabla_{\bm{e}_{\rm in}}\Phi
                + \bm{e}_{\rm in} \times \nabla_{\bm{j}_{\rm in}}\Phi}
            + \at{\rd{\bm{e}_{\rm in}}{t}}_{\rm GW},
            \label{eq:milankovich_e}\\
    \rd{\bm{L}_{\rm out}}{t}
        &= -\s{\bm{j}_{\rm out} \times \nabla_{\bm{j}_{\rm out}}\Phi
            + \bm{e}_{\rm out} \times \nabla_{\bm{e}_{\rm out}}\Phi},\\
    \rd{\bm{e}_{\rm out}}{t}
        &= -\frac{1}{L_{\rm out}}
            \s{\bm{j}_{\rm out} \times \nabla_{\bm{e}_{\rm out}}\Phi
                + \bm{e}_{\rm out} \times \nabla_{\bm{j}_{\rm out}}\Phi},
\end{align}
where the detailed forms of the $\Phi$ and the GW terms are well-known
\citep{peters1964, LL18, tremaine2023brown, grishin2024_brown} and are
reproduced in Appendix~\ref{app:eom_zlk} for simplicity.
Above, the $\bm{e}$ denote the two orbits' eccentricity vectors, $\bm{L}$ the
orbital angular momentum vectors (oriented along the orbit normal), and the
$\bm{j}$ are oriented along $\bm{L}$ and have magnitude $j = \sqrt{1 - e^2}$
(see Appendix~\ref{app:eom_zlk} for precise definitions).
We implement these gradients using the computer algebra system \texttt{sympy}
\citep{sympy}.

\begin{figure}
    \centering
    \includegraphics[width=\columnwidth]{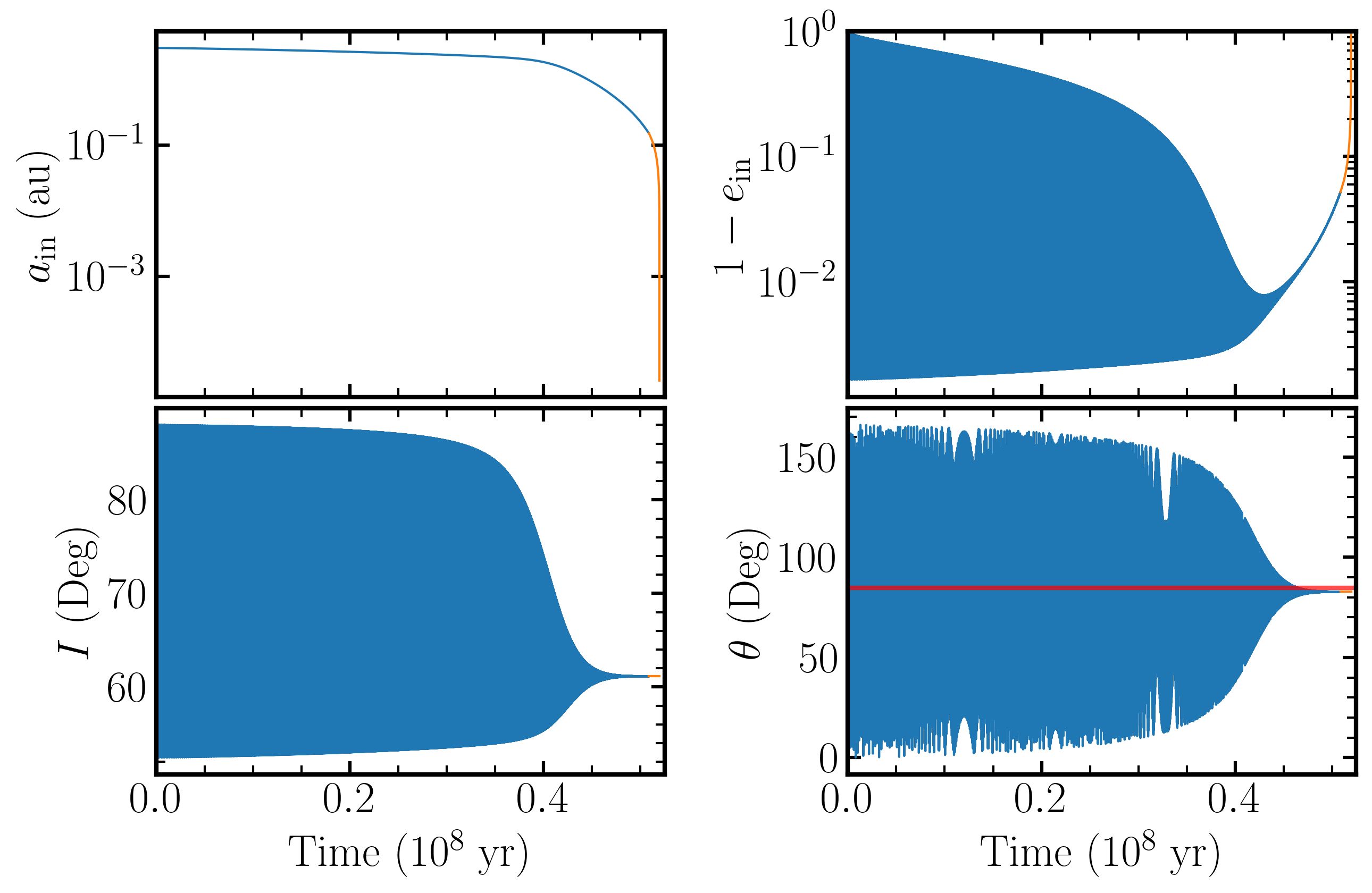}
    \caption{
    A binary's semi-major axis (top left), eccentricity (top right),
    inclination (bottom left), and spin-orbit misalignment angle $\theta$
    (bottom right) as it coalesces via the tertiary-induced merger channel in the
    vicinity of a SMBH\@.
    As the eccentricity and inclination oscillate periodically due to the ZLK
    effect, the enhanced emission of GW induces the BBH to merge.
    Note that $\theta$ experiences significant oscillations but eventually
    converges to the prediction of Eq.~\eqref{eq:naive} (horizontal red line).
    For computational efficiency, the late stages of inspiral are performed in
    the absence of the tertiary and without tracking the spin phase; this is
    denoted by the orange line in all four panels.
    Parameters used are: $a_{\rm in, 0} = 3\;\mathrm{au}$, $a_{\rm out} =
    8000\;\mathrm{au} = 0.04\;\mathrm{pc}$, $e_{\rm out} = 0.6$, $m_1 =
    33M_{\odot}$, $m_2 = 17M_{\odot}$, $m_3 = 10^7M_{\odot}$, and $I_0 =
    88^\circ$.
    }\label{fig:example}
\end{figure}

An example of this evolution is shown in Fig.~\ref{fig:example}, where the
enhanced GW radiation due to eccentricities generated by the ZLK effect are able
to drive the BBH to merge in $\lesssim 0.1\;\mathrm{Gyr}$ despite it being too
wide to merge in isolation (Eq.~\ref{eq:Tmerger}).
For computational efficiency, the full spin-orbit evolution is replaced with a
simplified evolution consisting of just GW emission at late times ($a_{\rm in}
\lesssim 0.2\;\mathrm{au}$) when the binary evolution is fully decoupled from
the effect of the SMBH\@; this is shown as the orange lines in all panels of
Fig.~\ref{fig:example}.
The evolution is truncated when $e_{\rm in} < 10^{-3}$;
note that the binary is still well outside the LVK frequency band (GW frequency
$10\;\mathrm{Hz}$, or $a_{\rm in} \sim 10^{-7}\;\mathrm{au}$) but will merge
shortly.

\subsubsection{Spin Evolution: An Adiabatic Invariant}\label{ss:ad_invar}

As the BBH gradually coalesces, the spins of the component black holes also
evolve.
At leading 1-PN order, spin-spin coupling can be neglected \citep{racine2008,
LL18}, so the evolution of the two BHs can be treated independently.
Thus, we will study the evolution of $\bm{S}_1 = S_1\uv{s}_1$ the spin of $m_1$,
though the evolution of $\bm{S}_2$ proceeds analogously.
For brevity, we will drop the subscript.
The de Sitter precession of $\uv{s}$ about $\jhat_{\rm in}$ is given by
\citep[e.g.,][]{barker1975, LL18}
\begin{align}
    \rd{\uv{s}}{t}
        &=
            \Omega_{\rm dS}
            \jhat_{\rm in} \times \uv{s},\label{eq:de_sitter}\\
    \Omega_{\rm dS}
        &=
            \frac{3G\p{m_2 + \mu_{\rm in}/3}}{2c^2a_{\rm in}j_{\rm
            in}^2}n_{\rm in}
            .
\end{align}
Note that the misalignment angle $\theta = \cos^{-1}\p{\uv{s} \cdot \jhat_{\rm
in}}$ is precisely the angle constrained by the GWTC-4.0 catalog.
The evolution of $\theta$ is shown in the bottom-right panel of
Fig.~\ref{fig:example} (see also \citealp{LL18, su2021_lk90}).
While $\theta$ appears to converge to its mean value by the end of the
evolution, such a behavior is not an obvious consequence of the spin evolution
given by Eq.~\eqref{eq:de_sitter}: there is no dissipation!

Instead, the mechanism for this convergence is due to an unexpected adiabatic
invariant (\citealp{LL18, su2021_lk90}; see also \citealp{yu2020_spin}).
Given that the analytical origins of this mechanism were previously
derived in extensive detail in our previous work \citep{su2021_lk90}, we omit a
full rederivation here.
For interested readers, we provide an alternative derivation with a greater
physical emphasis in Appendix~\ref{app:spin_adinvar}.
For the purposes of this paper, the key result is as follows:
if a binary black hole is formed with spin-orbit alignment
and undergoes a sufficiently adiabatic (merger time $\gg$ the ZLK cycle time)
merger, then the final
spin-orbit misalignment angle $\theta$ is given by:
\begin{align}
    \bar{\mathcal{A}}_0
        \equiv{}&
            \frac{3Gm_{12}(m_2 + \mu_{\rm in} / 3) a_{\rm out}^3j_{\rm out}^3}{
                2c^2m_3a_{\rm in, 0}^4}\nonumber\\
            &\times \tilde{\mathcal{A}}(a_{\rm in, 0}, I_0,\dots)
            ,\\
    \cos \theta_{\rm f}
        \simeq{}& \abs{\cos I_0 - \bar{\mathcal{A}}_0},\label{eq:naive}
\end{align}
where $\tilde{\mathcal{A}}$ is some dimensionless function of order unity (that
primarily depends on the detailed value of $e_{\rm in, \max}$).
The accuracy of Eq.~\eqref{eq:naive} is illustrated as the horizontal red line
in the bottom-right panel of Fig.~\ref{fig:example}.
We see that in the limit $\bar{\mathcal{A}}_0 \to 0$ and $I_0 \approx 90^\circ$
that we recover the $90^\circ$ attractor result.
However, for more compact binaries, $\mathcal{A}$ is not so small, and a
positive bias in the observed $\cos \theta$ distribution arises, consistent with the
LVK constraints (as discussed in Section~\ref{s:intro}).

\begin{figure}
    \centering
    \includegraphics[width=0.7\columnwidth]{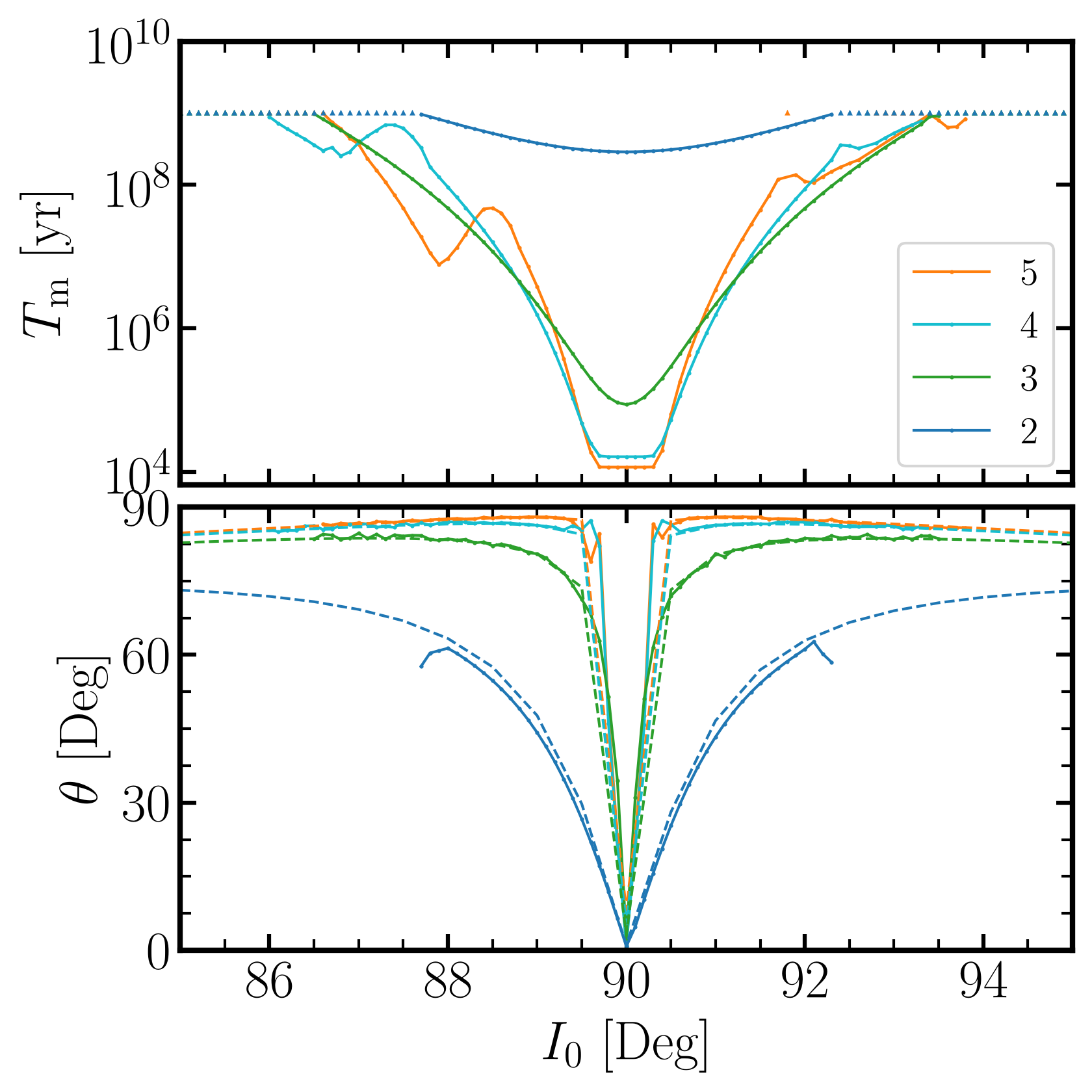}
    \includegraphics[width=0.7\columnwidth]{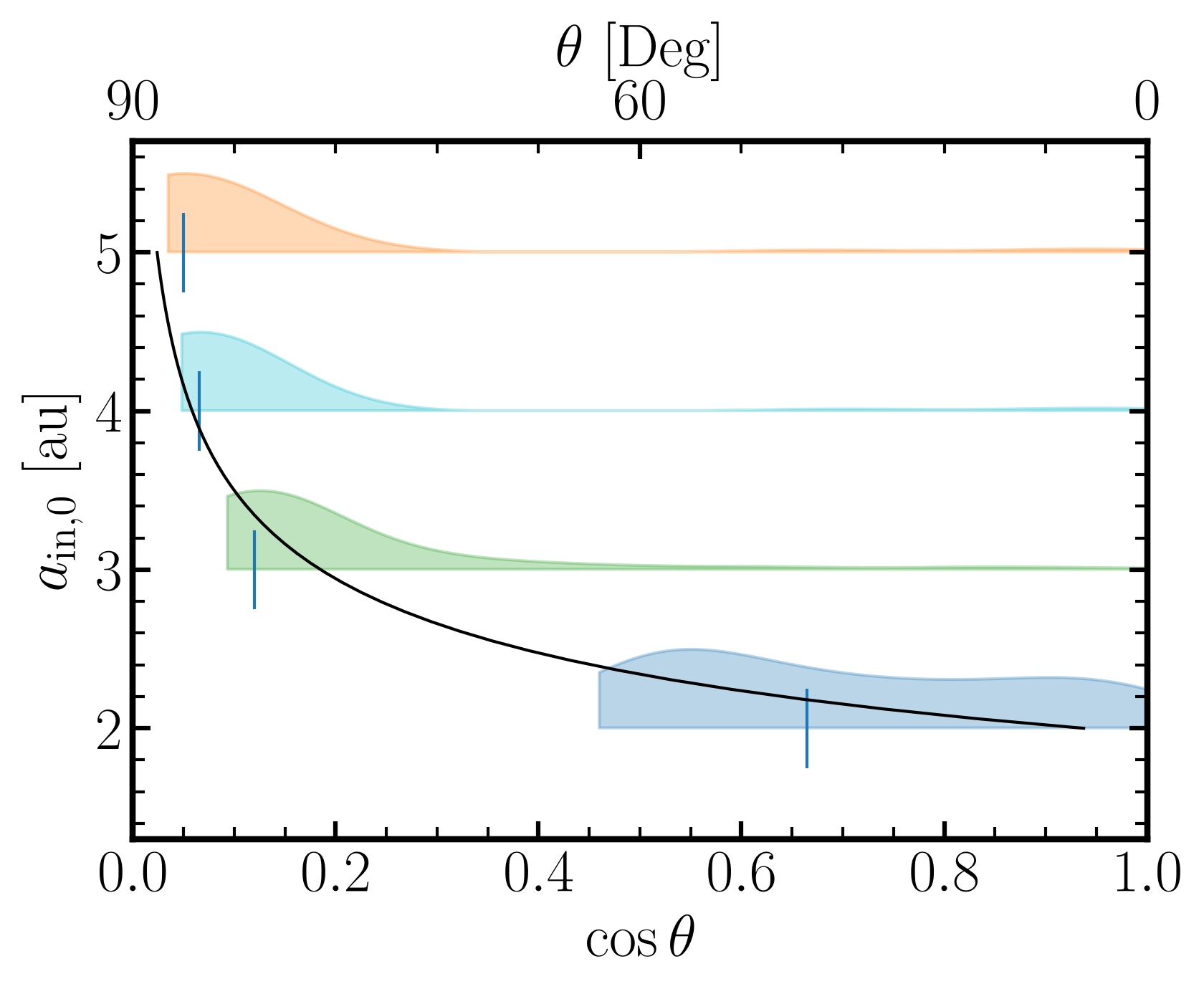}
    \caption{
    In these plots, we show how the final spin orientations of tertiary-induced
    BBH mergers change with their initial semi-major axis (we choose to vary
    $a_{\rm in, 0} \in \z{2, 3, 4, 5}\;\mathrm{au}$) across a range of initial
    mutual inclinations $I_0$ between the inner and outer orbits.
    In the top panel, we show the merger times as a function of $I_0$ and
    $a_{\rm in, 0}$ (legend), where
    systems that evolve for longer than $1\;\mathrm{Gyr}$ are labeled with
    triangles (and likely become unbound by relaxation processes in the NSC).
    In the middle panel, we show the resulting misalignment angles for all
    merging systems.
    For each value of $a_{\rm in, 0}$, we also show the numerical evaluation of
    Eq.~\eqref{eq:naive} in the correspondingly colored dashed lines; good
    agreement is seen.
    Finally, in the bottom panel, we show the histograms of $\cos \theta$
    obtained at each value of $a_{\rm in, 0}$, and we have overlaid an
    arbitrary $a_{\rm in, 0}^{-4}$ line (predicted by Eq.~\ref{eq:naive}) to
    guide the eye.
    All parameters other than those labeled are the same as in
    Fig.~\ref{fig:example}.
    }\label{fig:violin_bbhonly}
\end{figure}

We compare these analytical results to direct numerical integrations performed
with a few values of $a_{\rm in, 0} \in \z{2, 3, 4, 5}\;\mathrm{au}$ in
Fig.~\ref{fig:violin_bbhonly}.
In the top panel, we show the merger times of the BBH systems that successfully
merge within the duration of the numerical integration, $1\;\mathrm{Gyr}$.
In the middle panel, we show that the resulting misalignment angles can be
well-described by Eq.~\eqref{eq:naive}.
The central dip is because $\Omega_{\rm ZLK} \to 0$ as $I \to 90^\circ$, and the
effective precession axis is aligned with $\jhat_{\rm in}$ at all times
\citep{su2021_lk90}.
Other deviations from the analytical result likely arise from non-adiabaticity
or resonances, both of which can be obtained from a more careful analysis than
presented here \citep{su2021_lk90}.
Finally, in the bottom panel, we show the spin-orbit misalignment angle
distributions at each $a_{\rm in, 0}$.
The distributions follow the naive $a_{\rm in, 0}^{-4}$ predicted by
Eq.~\eqref{eq:naive} quite well.
Small deviations arise when $\bar{\mathcal{A}}_0 \lesssim \cos I_0$, which
violates the approximation used to derive Eq.~\eqref{eq:naive}.

It is important to note that the anti-correlation between $a_{\rm in, 0}$ and
$\cos \theta$, a result of Eq.~\eqref{eq:naive}, relies on the assumption that
$\uv{s}_0 \parallel \jhat_{\rm in, 0}$, a limiting assumption that was also
pointed out in \citet{yu2020_spin}.
For dynamically-assembled BBHs, such an assumption cannot be made, as the
component BH spins are randomly oriented upon capture.
For BBHs born from wide stellar binaries, there are two issues with this
initial condition:
the spins of the stars need not be aligned with the orbits of their wide binary
companions at birth, and the birth of the BHs will significantly modify the
orbit due to supernova (SN) kicks \citep[e.g.,][]{rodriguez2016binary}.
On the other hand, for more compact BBHs, this is a sensible initial condition
because the Keplerian orbital velocity of the binary becomes larger than the
natal kick velocity, and because the stellar spins are likely well-aligned
with the orbit (whether because the stars are co-natal or due to alignment
processes such as mass transfer).

On its own, a correlation between $a_{\rm in, 0}$ and $\theta$ is not so
useful, as $a_{\rm in, 0}$ is a property of the BBH at its formation rather than
when it is observed by LVK\@.
However, if $a_{\rm in, 0}$ is correlated with other properties of the BBH,
then the mechanism we've discussed can introduce additional structures to the
observed parameters of BBH systems.
In the next section, we show that a preceding phase of binary stellar evolution
can introduce such correlations.

\subsection{Stellar Binary Evolution}\label{s:stellar_evol}

As the origin of the BBHs considered in the previous section, we consider that
they may be the final evolutionary stage of stellar binaries orbiting a SMBH\@.
Observationally, stellar binaries have been found within $\sim 0.1\;\mathrm{pc}$
of Sgr A$^*$ \citep{martins2006_sga1, pfuhl2014_sga2, peissker2024_binarySgr},
consistent with the binary fraction of young star clusters
\citep{alexander2017_nscreview}.
On the other hand, more recent studies identify a deficit of massive
stellar binaries within the central $0.02\;\mathrm{pc}$ of Sgr A$^*$
\citep{chu2023_sgrabin}.
Nevertheless, as the formation of nuclear star clusters (NSCs) is not well understood
\citep[see e.g.][for a recent review]{neumayer2020}, it appears plausible that
massive stellar binaries can be formed as close in as $\sim 0.04\;\mathrm{pc}$
to a central SMBH (corresponding to our fiducial parameters, see
Fig.~\ref{fig:example}) on eccentric orbits \citep{ali2020_nscthermal}.
Additionally, due to the low efficiency of dynamical binary formation so near an
SMBH, where the velocity dispersion is very large \citep[e.g.,][]{hut1985nsc,
quinlan1989binary}, it is likely that any BBHs so close to their central SMBHs
either formed as a primordial stellar binary or formed via gas-assisted capture
in an active galactic nucleus (AGN) disk \citep[e.g.,][]{bartos2017}.
In this section, we focus on the former possibility and defer discussion of the
latter to many other excellent works (see Section~\ref{s:discussion}).

In this section, we consider an initial stellar binary with total mass $m_{12,
\star} = 100M_{\odot}$ at semi-major axis $a_{\rm in,\star} = 2\;\mathrm{au}$,
and with the same outer orbital parameters used above in
Section~\ref{ss:ad_invar}.
We will study the resulting spin misalignment of the merging BBH as a function
of the BBH mass ratio $q_{\rm BH}$ by varying the mass ratio of the initial
stellar binary $q_\star$.

\subsubsection{Short-Range Force Suppression of ZLK}\label{ss:bse_srf}

First, we address one key ingredient for this mechanism.
In order for the $a_{\rm in, \star} \sim \mathrm{au}$ stellar binary to form a
$a_{\rm in} \sim \mathrm{au}$ BBH that subsequently merges, two conditions must
be met: the binary must not experience ZLK-driven coalescence on the stellar
phase, but the subsequent BBH must reach sufficiently large eccentricities via
ZLK to be driven towards merger.
One way that this can be accomplished is by reorienting the binary between its
stellar and BH phases, either via collective effects such as resonant relaxation
\citep{rauch1996_vrr} or via natal kicks \citep[e.g.,][]{vignagomez2023natal}.
We consider a second possibility, that the additional sources of apsidal
precession present in stellar binaries (often simply referred to as short-range
forces, or SRFs) are able to suppress ZLK \citep[e.g.,][]{kiseleva1998_triples,
eggleton2001, wu2003_HJ, liu2015suppression} while those in the subsequent BBH
cannot.
Since these results are well-established, we defer their discussion to
Appendix~\ref{app:SRF} and quote the key results here.

For two stars on sufficiently wide orbits that their spins do not tidally
synchronize on their main sequence lifetimes ($a_{\rm in} \gtrsim 1
\;\mathrm{au}$), the dominant source of apsidal precession is due to the
non-Newtonian potential generated by their rotational bulges. This precession
can suppress ZLK oscillations if the sums of the dimensionless parameters
$\epsilon_{\rm rot, 1} + \epsilon_{\rm rot, 2} > 9/4$
\citep[e.g.,][]{liu2015suppression}, where
\begin{align}
    \epsilon_{\rm rot, 1}
        ={}& 2.3 \frac{m_{1,\star}}{m_{2,\star}}
            \p{\frac{k_{q, 1}}{0.02}}
            \p{\frac{m_{12,\star}}{100M_\odot}}
            \p{\frac{\tilde{a}_{\rm out}}{6400\;\mathrm{au}}}^3
                \nonumber\\
        &\times \p{\frac{R_1}{10 R_{\odot}}}^2
            \p{\frac{a_{\rm in, \star}}{2\mathrm{au}}}^{-5}
            \p{\frac{m_3}{10^7M_{\odot}}}^{-1}\nonumber\\
        &\times \p{\frac{\Omega_{\rm s, 1} / \Omega_{\rm dyn, 1}}{0.8}}^2.
            \label{eq:epsrot_val}
\end{align}
On the other hand, in binary black hole systems, the dominant source of apsidal
precession is general relativistic pericenter advance, a first-order
post-Newtonian effect.
This effect will suppress ZLK oscillations if the parameter $\epsilon_{\rm GR} >
9/4$ \citep{liu2015suppression}, where
\begin{align}
    \epsilon_{\rm GR}
        ={}& 0.12
            \p{\frac{m_{\rm 12}}{50M_\odot}}^2
            \p{\frac{\tilde{a}_{\rm out}}{6400\;\mathrm{au}}}^3\nonumber\\
        &\times \p{\frac{a_{\rm in}}{2\mathrm{au}}}^{-4}
            \p{\frac{m_3}{10^7M_{\odot}}}^{-1}.\label{eq:eps_gr}
\end{align}

\subsubsection{Binary Stellar Evolution: Double Stable Mass Transfer}\label{ss:binary_evol}

Given that the stellar binary, consisting of two stars with zero-age main
sequence (ZAMS) masses $m_{1, \star}$ and $m_{2, \star}$, does not experience ZLK
oscillations, it will evolve according to standard theories of isolated binary
evolution.
In traditional prescriptions, such evolution results in one phase of stable mass
transfer followed by a second phase of unstable mass transfer (a common envelope
phase), leading to a very compact binary that may even successfully merge in
isolation \citep[e.g.,][]{hurley2002, belczynski2016first}.
However, more recent studies suggest that, except for very extreme mass ratios
($q_\star \equiv m_{2,\star} / m_{1,\star} \lesssim 0.2$), many stellar binaries
that form compact object binaries may instead undergo two phases of stable mass
transfer (MT) \citep[e.g.,][]{gallegosgarcia2021_2mt, vanson2022_analmt}.
As the detailed physics of these two phases of MT are still filled
with uncertainties, we can gain some qualitative understanding of the parameter
space that can result by implementing simple analytical prescriptions.
In general, the final outcome of MT will depend on the stellar
properties, and will introduce correlations between the initial $a_{\rm in}$ of
the BBH and its various other physical properties.
We describe the double MT phase using the formalism of \citet{soberman1997}
(S97) and use similar parameters as modern works
\citep[e.g.,][]{vanson2022_analmt, riley2022_compas}.

In the S97 formalism, as a star $m_{1,\star}$ initiates MT, the mass
lost from the star follows one of three modes.
The first is called Jeans's mode or the fast mode \citep{soberman1997}, in which
a spherically symmetric outflow directly removes matter from $m_{1,\star}$,
carrying away the specific angular momentum of the donor star.
The second is called isotropic re-emission, in which matter is first transferred
to the vicinity of the accretor before being rapidly, isotropically ejected;
matter lost in this fashion carries away the specific angular momentum of the
accretor instead.
The third is accretion, in which the matter lost from $m_{1,\star}$ is deposited
onto $m_{2,\star}$, conserving the total angular momentum of the system.
A given MT phase can be parameterized by the fractions $\alpha$,
$\beta$, and $\epsilon$, which denote the fraction of matter lost from the
primary via either the fast mode, isotropic re-emission, or accretion
respectively; note that $\alpha + \beta + \epsilon = 1$ (we do not consider mass
loss through L2, which becomes important for rapid mass loss,
\citealp{soberman1997, lu2023_MTL2}).

In this work, we will assume that, between the zero-age main sequence and the
end of the mass transfer phase, the primary loses the entirety of its radiative
envelope, and we denote $f_{\rm core}$ to be the core mass fraction of the
primary \citep[see e.g.][]{vanson2022_analmt}.
For the high-mass stars we consider, $f_{\rm core} \simeq 0.5$ (broadly
consistent with the MIST stellar evolution tracks, \citealp{dotter2016mist1,
choi2016mist2}), and as a first approximation we take this to be a constant; we
will explore a mass-dependent $f_{\rm core}$ shortly.
Then, the evolution from a ZAMS binary to a BBH is modeled in four separate
stages:
\begin{itemize}
    \item During the first phase, the primary ejects some fraction $\eta_{\rm
        wind}$ of its envelope mass, $(1 - f_{\rm core})m_{1,\star}$, due to
        stellar winds.
        The exact value of $\eta_{\rm wind}$ is currently not well-known for
        massive stars, but it is likely subdominant to binary-driven mass loss
        \citep{smith2014_masslossrev, vinciguerra2020_MT}, implying that
        $\eta_{\rm wind} \lesssim 0.5$.
        We take $\eta_{\rm wind} = 0.2$ as a fiducial value.

        Among this wind-ejected matter, we approximate that half will be
        unbound from the binary, and the other half will be accreted by the
        secondary \citep[as seen in high-mass x-ray binaries,
        e.g.][]{fornasini2024high}.
        Thus, we take $(\alpha_1, \beta_1, \epsilon_1) = (0.5, 0, 0.5)$ as
        fiducial values for this wind-driven mass loss phase.

    \item During the second phase, the primary ejects the remainder of its
        envelope (consisting of mass $(1 - \eta_{\rm wind})(1 - f_{\rm core})m_{1,
        \star}$) on a thermal timescale.
        As such, the fast mode of mass loss is sufficiently weak to be
        neglected \citep[e.g.,][]{vanson2022_analmt}.
        Of the remaining mass, all of which reaches the vicinity of the
        secondary, some fails to be accreted and experiences isotropic
        re-emission.

        For simplicity, let us define $\tilde{\beta}_2 \equiv \beta_2 / (1 -
        \eta_{\rm wind})$, and similarly for $\tilde{\epsilon}_2$.
        As such, the three quantities satisfying
        \begin{equation}
            \eta_{\rm wind} +
            \tilde{\beta}_2 + \tilde{\epsilon}_2 = 1,\label{eq:fracs_tilde}
        \end{equation}
        denote the mass fractions of the primary's ejected envelope lost via
        stellar winds, isotropic re-emission during stable MT, and accretion
        during stable MT respectively.
        To account for this, we take fiducial values $(\tilde{\beta}_2,
        \tilde{\epsilon}_2) = (0.4, 0.4)$ for this mass transfer phase.
        This is implemented taking $(\alpha, \beta, \epsilon) = (0, 1/2, 1/2)$.

        After this phase, we model that the remaining mass of the primary,
        $f_{\rm core}m_{1,\star}$ experiences direct collapse to a BH with no
        natal kick \citep[e.g.,][]{Giacobbo2020, de2024_directcollapse}.
        The secondary has attained a new mass $m_{2, \star} = m_{2, \star,
        \mathrm{ZAMS}} + \tilde{\epsilon}_2(1 - f_{\rm core})m_{1,\star}$.
        Its new core mass is then $f_{\rm core}m_{2, \star}$, i.e.\ the
        convective-radiative boundary of the accretor adjusts to its new mass
        (and possibly ``rejuvenating'' the core, e.g.\
        \citealp{hellings1983rejuv, renzo2021_accretion}, but see also
        \citealp{braun1995_rejuv}).

    \item During the third phase, the secondary also experiences wind-driven
        mass loss.
        It also loses $\eta_{\rm wind}$ of its envelope, with $(\alpha_3,
        \beta_3, \epsilon_3) = (0.5, 0, 0.5)$.

    \item During the fourth phase, the secondary initiates a second phase of
        stable MT\@.
        Again, the fast mode is too slow to contribute to mass loss.
        However, additionally, the primary (which is now a BH) likely does not
        accrete efficiently, i.e.\ not much more rapidly than the strongly
        constraining Eddington limit \citep{vanson2022_analmt}; note that even
        sustained accretion at $\sim 10^2L_{\rm edd}$, as suggested by recent 3D
        radiation hydrodynamics simulations \citep{toyouchi2024_rhdsuperedd},
        only leads to modest accretion during a stable MT episode.
        Due to the low accretion efficiency, almost all of the mass transferred
        from the secondary experiences isotropic re-emission.
        As such, we take $\tilde{\beta}_4 = 1$ and $\tilde{\epsilon}_4 = 0$ as
        fiducial values.
\end{itemize}

In summary, the four phases of the mass transfer procedure depend on a total of
five free parameters: $\eta_{\rm wind}$, $\tilde{\epsilon}_2$,
$\tilde{\epsilon}_4$, and the two $f_{\rm core}$ values for the stars (recall
that $\tilde{\beta}_2 = 1 - \eta_{\rm wind} - \tilde{\epsilon}_{\rm 2}$ and
similarly for $\tilde{\beta}_4$).
By fixing $\tilde{\epsilon}_4 = 0$ and enforcing a shared $f_{\rm core}$
prescription for the two stars (either constant or mass-dependent), the number
of free parameters is reduced to just three.

Regarding the spin evolution, we assume that the binary stars form with
exact spin-orbit alignment (which is expected at such close separations, though
there may be evidence for misalignment in eccentric binaries,
\citealp{marcussen2024_ecc}).
Then, subsequent evolution does not change the orbit orientation (natal kicks are
neglected), and the spin orientation also does not change (the secondary
is aligned by accretion from the primary during the second MT phase, while the
primary never accretes much mass during the fourth MT phase and retains its
primordial orientation).

\begin{figure}
    \centering
    \makebox[\columnwidth][r]{%
        \includegraphics[width=0.55\columnwidth]{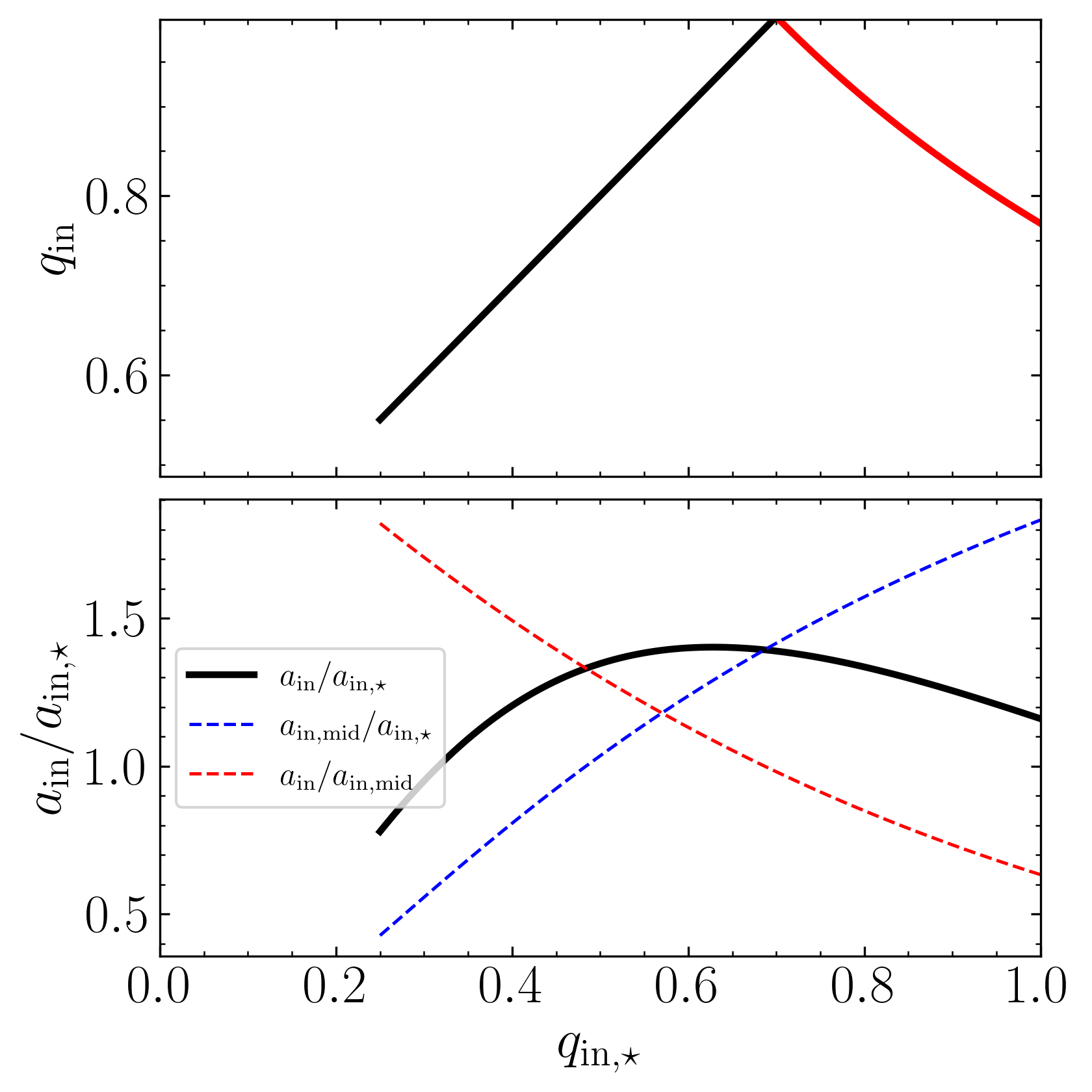}%
        \hspace{0.02\columnwidth}%
        \includegraphics[width=0.55\columnwidth]{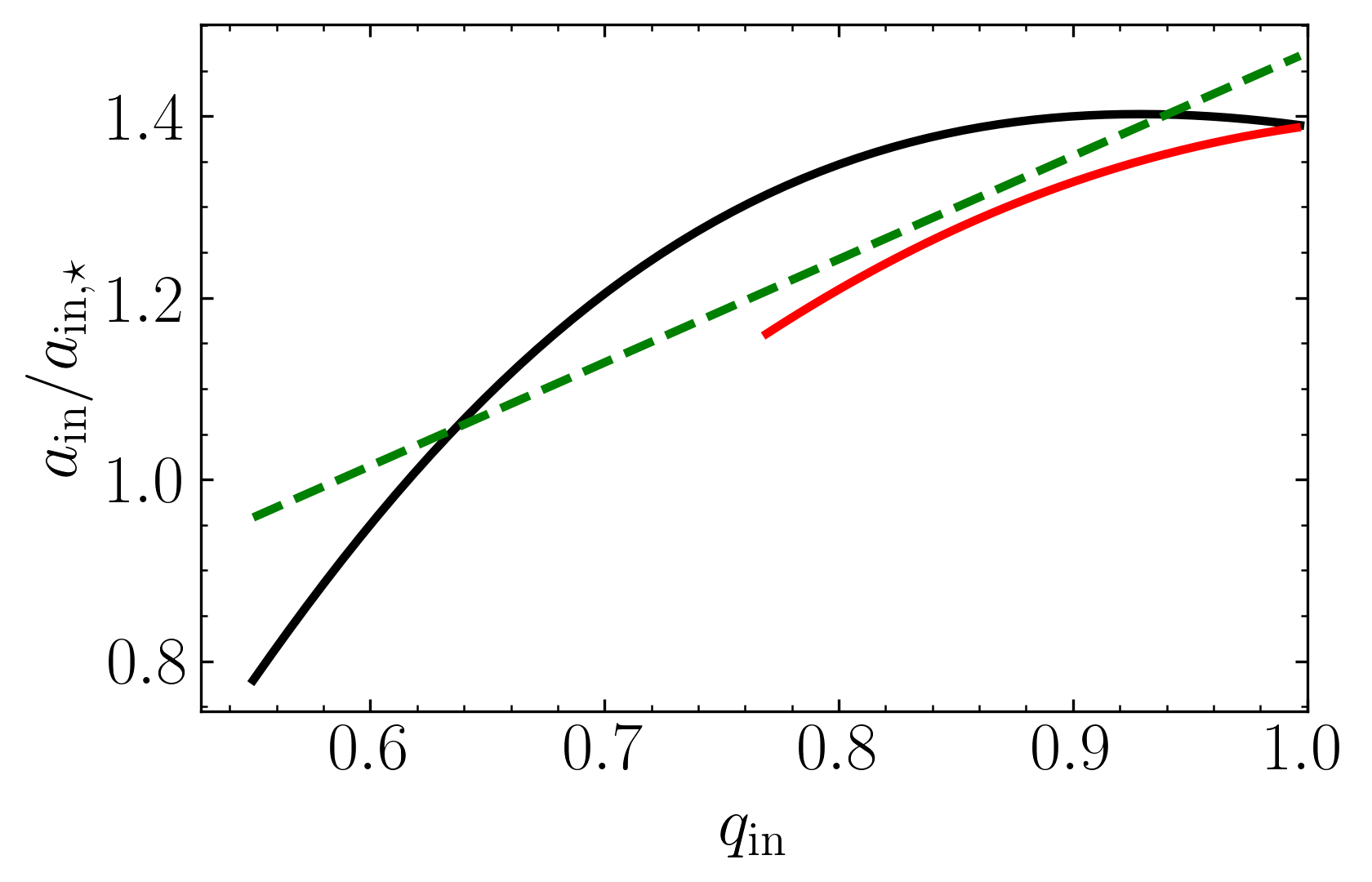}%
    }
    \caption{
    In the top-left panel, we show the final mass ratio $q_{\rm in}$ (i.e., that
    of the binary black hole) as a function of $q_{\rm in, \star}$ obtained
    after the four-step mass transfer prescription discussed in
    Section~\ref{ss:binary_evol}.
    The red line denotes systems that experience mass ratio inversion, where the
    more massive BH is formed from the initially less massive star, and the
    black line denotes systems that do not experience this inversion.
    In the bottom-left panel, we illustrate the three pairwise ratios of the binary's
    initial semi-major axis $a_{\rm in, \star}$, its intermediate value (after
    the formation of the first BH, i.e.\ at the end of the second phase) $a_{\rm
    in, mid}$, and its final value (after the formation of both BHs) $a_{\rm
    in}$ also as functions of $q_{\rm in, \star}$.
    In the right panel, we show the resulting correlation between $a_{\rm in}$
    and $q_{\rm in}$, where the black and red lines have the same meaning as in
    the top panel.
    A positive correlation is seen, as depicted by the linear fit shown as the
    green dashed line.
    The parameters used here are $f_{\rm core} = 0.5$, $\eta_{\rm wind} = 0.2$,
    $\tilde{\epsilon}_2 = 0.4$, and $\tilde{\beta}_2 = 0.4$.
    }\label{fig:mt_res}
\end{figure}

By applying this four-phase prescription for mass transfer, we obtain a relation
between the stellar binary's semi-major axis $a_{\rm in, \star}$ and that of
the BBH $a_{\rm in}$ (for consistency with the notation
in Section~\ref{s:spindyn}, we omit the BH subscripts for quantities describing
the BBH) as well as between the stellar binary's mass ratio $q_{\rm in, \star}$
and that of the BBH $q_{\rm in}$.
In Fig.~\ref{fig:mt_res}, we show the binary evolution resulting from the
four-phase MT prescription detailed above.
First, we note in the top panel that mass ratio inversions are seen for
near-equal-mass binaries (red line).
Since the spins of the first and second-formed BHs are sometimes expected to
differ \citep[e.g.,][]{olejak2024_unequalmass}, we will keep track of mass ratio
inversions.
Note that the MT prescription we've proposed is scale-free and only yields
predictions for the semi-major axis ratios rather than their specific values.
Of course, physical models of MT yield dramatically different results depending
on binary orbital separation, which would be modeled in our formalism as
dependencies of $\tilde{\epsilon}_2$ and other parameters on $a_{\rm in,
\star}$.

There are two features about the evolution above that are essential to our
subsequent discussion.
First, note that $a_{\rm in} \simeq a_{\rm in, \star}$, i.e.\ the binary does
not shrink appreciably during its evolution to a BBH, and GR precession does
not suppress ZLK oscillations.
Second, the last panel of Fig.~\ref{fig:mt_res} shows that $a_{\rm in}$ and
$q_{\rm in}$ are positively correlated.
Then, application of Eq.~\eqref{eq:naive} suggests that the final $\cos \theta$
will be \emph{negatively} correlated with $q_{\rm in}$; this will be shown in
Section~\ref{s:signatures}.

\begin{figure*}
    \centering
    \includegraphics[width=0.32\textwidth]{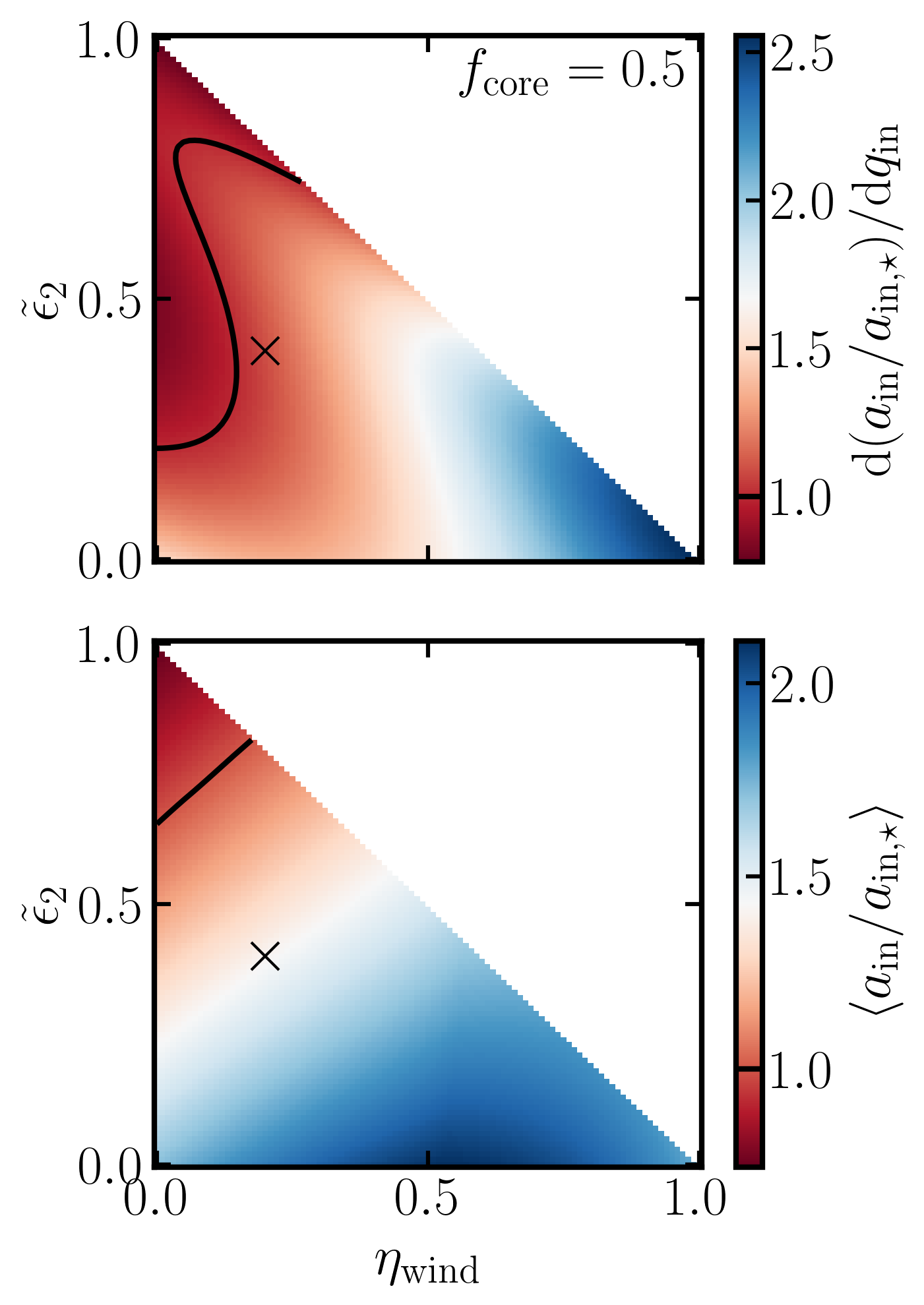}
    \includegraphics[width=0.32\textwidth]{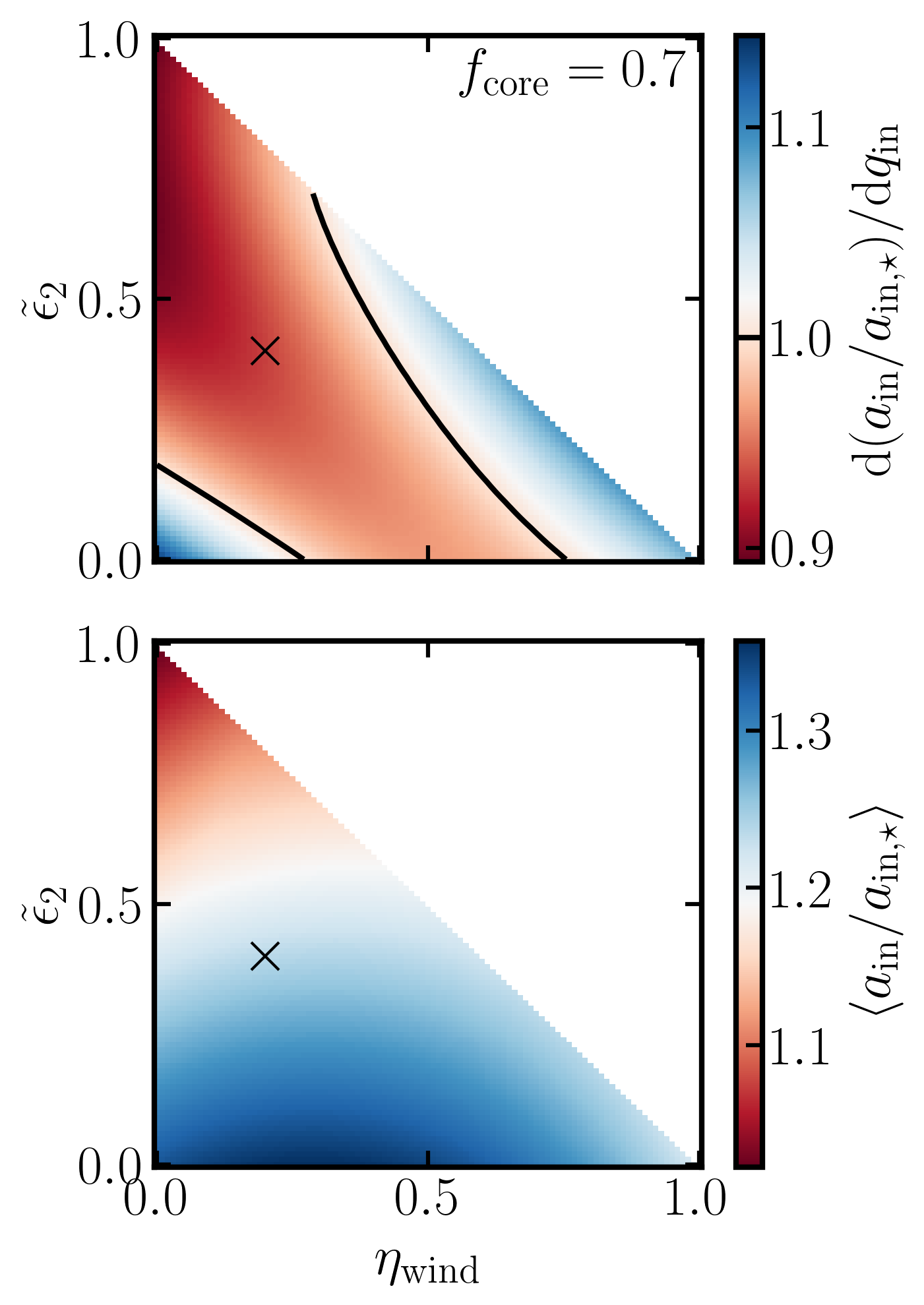}
    \includegraphics[width=0.32\textwidth]{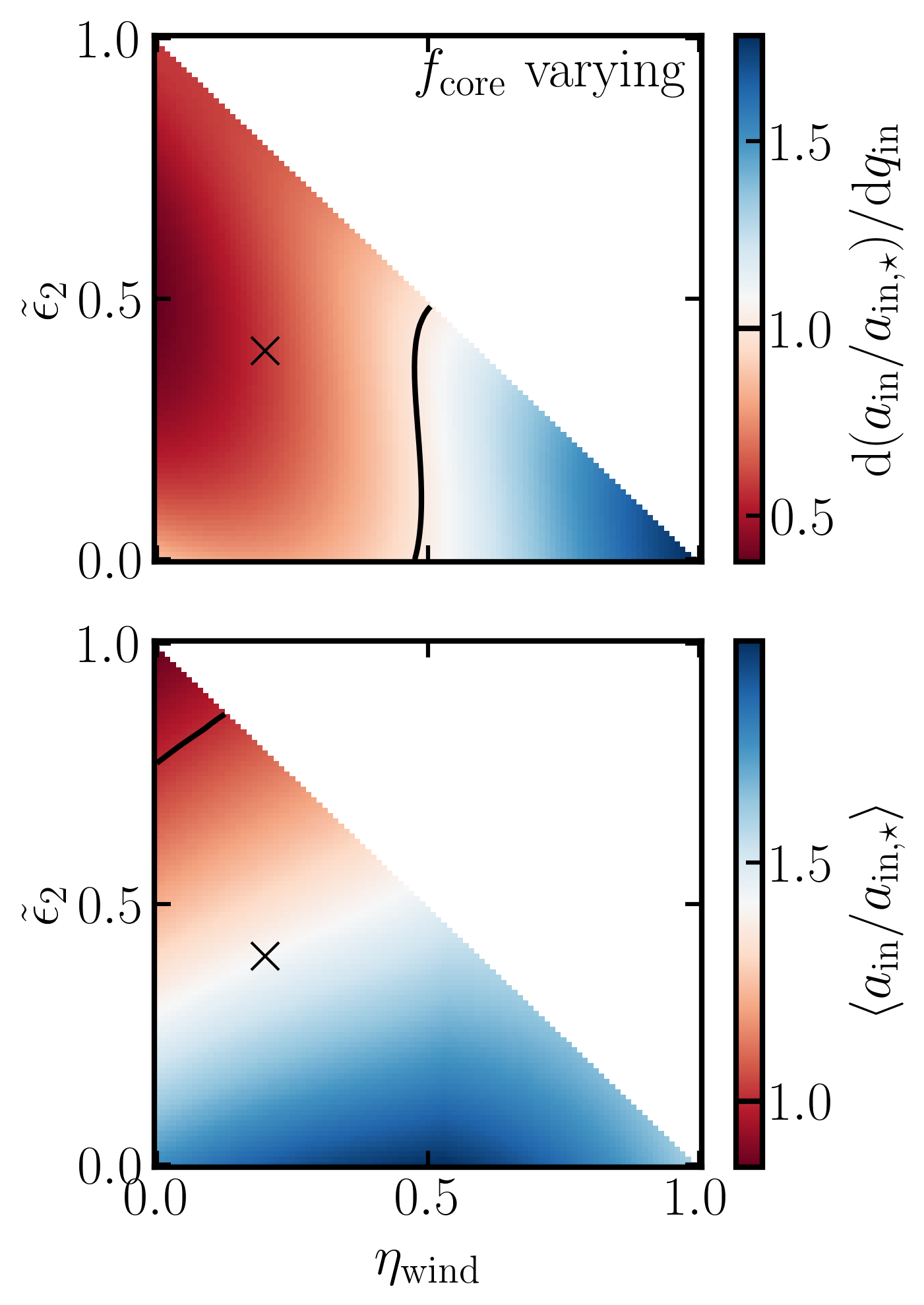}
    \caption{
    The left column of plots shows the values of the mean correlation
    $\rdil{(a_{\rm in} / a_{\rm in, \star})}{q_{\rm in}}$ and the mean semi-major axis change
    $\ev{a_{\rm in} / a_{\rm in, \star}}$ as a function of $\eta_{\rm wind}$ and
    $\tilde{\epsilon}_{\rm 2}$, two of the three free parameters of the MT
    prescription laid out in Section~\ref{ss:binary_evol}; the third, $f_{\rm
    core}$, is fixed at $0.5$.
    The upper-right triangle in both plots is forbidden due to the requirement
    that $\eta_{\rm wind} + \tilde{\epsilon}_2 + \tilde{\beta}_2 =
    1$ (Eq.~\ref{eq:fracs_tilde}) and that $\tilde{\beta}_2 \geq 0$.
    The middle column of plots shows the same but for $f_{\rm core} = 0.7$.
    The final column of plots shows the same but for a mass-dependent $f_{\rm
    core}$ (Eq.~\ref{eq:fcore_var}).
    In all panels, a value of $1$ is shown as a black contour when within the
    scale of the color plot.
    }\label{fig:98slopes}
\end{figure*}

To understand the robustness of these two features as a function of the three
free parameters of the system, we study the dependence of the mean slope
$\rdil{a_{\rm in}}{q_{\rm in}}$ (defined as the best fit line as shown in
Fig.~\ref{fig:mt_res}) and the mean semi-major axis change $\ev{a_{\rm in} /
a_{\rm in, \star}}$ on the free parameters of the MT prescription.
In Fig.~\ref{fig:98slopes}, we show the dependence of these two quantities on
$\eta_{\rm wind}$, $\tilde{\epsilon}_{2}$ for three different $f_{\rm core}$
prescriptions: $f_{\rm core} = 0.5$, $f_{\rm core} = 0.7$, and a mass-dependent
fraction
\begin{equation}
    f_{\rm core} = 0.4 + (m_\star / 320M_{\odot}).\label{eq:fcore_var}
\end{equation}
This simple prescription is in coarse agreement with asteroseismic and
spectroscopic observations for lower masses \citep{tkachenko2020_ccore,
johnstton2021_ccore, pederson2021_ccore} and with the MIST evolutionary tracks
at higher masses \citep{dotter2016mist1, choi2016mist2}.
Broadly, it can be seen that a positive correlation between $a_{\rm in}$ and
$q_{\rm in}$, as well as a gentle orbital softening, are robust features of our
MT model.

As a final note, we comment on the intuitive reason for the observed correlation
between $a_{\rm in}$ and $q_{\rm in}$ during our double MT prescription.
First, conservative mass transfer generally drives small-mass-ratio systems
closer to equal mass ratios.
This effect is preferentially stronger for more unequal initial mass ratios.
Since conservative mass transfer conserves $L_{\rm in} \propto q_{\rm in, \star}
\sqrt{a_{\rm in, \star}} / (1 + q_{\rm in, \star})$, it can be seen that a
larger increase in $q_{\rm in, \star}$ corresponds to a larger decrease of
$a_{\rm in, \star}$.
Accordingly, we find that BBHs formed with small $q_{\rm in}$ should also have
small $a_{\rm in}$.
Second, isotropic re-emission leads to a similar correlation: since the ejected
matter leaves with the specific angular momentum of the accretor, a smaller mass
ratio ejects more angular momentum and results in more binary hardening.
Finally, fast mode mass loss has a smaller effect on the orbital separation
(since it is typically ejected from the more massive donor, and as such it
carries away little angular momentum) but broadly leads to binary widening.
Thus, the combination of all three mass loss mechanisms results in
a combination of mass-ratio-dependent binary hardening and general wind-driven
binary softening.

\section{Results}\label{s:signatures}

As shown in the previous section, a population of stellar binaries with fixed
semi-major axis $a_{\rm in, \star}$ and varying mass ratios $q_{\rm in, \star}$
may yield a population of BBHs with positively correlated $a_{\rm in}$ and
$q_{\rm in}$.
Such a result can then produce an anti-correlation between $\cos \theta$ and
$q_{\rm in}$
following the results of Section~\ref{ss:ad_invar}, namely Eq.~\eqref{eq:naive}.
In this section, we use numerical integrations to explore the strength of this
correlation and substantiate the analytical arguments made above.

For clarity, we consider a population of nearly-circular ($e_{\rm in} =
10^{-3}$) stellar binaries with fixed total mass $m_{12, \star} = 100M_{\odot}$,
fixed semi-major axis $a_{\rm in, \star}$, and with $q_{\rm in, \star} \in
[0.25, 1]$ \citep[a standard MT stability threshold, e.g.][]{schneider2015_mt,
temmink2023_MTstab}.
Using the results of the previous section, these stellar binaries evolve into
BBHs via the MT prescription of Section~\ref{ss:binary_evol} with semi-major
axes $a_{\rm in}$ and mass ratios $q_{\rm in}$.
Then, for the orbital dynamics phase of evolution, these BBHs are assumed to
have a random initial inclination with respect to the plane of their orbit around a
central SMBH (though for computational efficiency we restrict $I$ to values
within the window of inclinations that can drive successful mergers as shown in
the upper panels of Fig.~\ref{fig:violin_bbhonly}).
The remaining orbital elements are held fixed for simplicity: $\omega_{\rm in} =
0$, $\omega_{\rm out} = 0.7$, and $\Omega_{\rm in} = \Omega_{\rm out} - \pi = 0$.
Finally, the initial spins of the BHs are aligned with the inner orbit normal,
due to the multiple phases of MT\@.
The inner and outer orbits and the spins of the BHs are then evolved following
the equations of motion as described in Section~\ref{s:spindyn} for
$1\;\mathrm{Gyr}$, which is significantly longer than the expected unbinding time
for such binaries in typical NSC models (see Section~\ref{s:signatures}).
For systems that successfully merge, we record their merger time and final
spin-orbit misalignment angle.

Note that we must account for the possibility that the rotational SRF
(Eq.~\ref{eq:epsrot_val}) may be too weak to suppress ZLK excitation after the
formation of the first BH\@.
To do so, we evaluate Eq.~\eqref{eq:epsrot_val} at the end of
the second MT phase (i.e.\ using $a_{\rm in, mid}$ from Fig.~\ref{fig:mt_res})
but, for simplicity, we hold the stellar parameters $R_\star$ and $k_{\rm q}$
constant at their fiducial values (which avoids having to introduce detailed
stellar modeling, beyond the scope of this paper), and we set $\Omega_{\rm s} /
\Omega_{\rm dyn} = 1$ for the secondary (which is spun up to near critical
rotation by the MT from the primary).
We then keep track of systems that pass through a state with $\epsilon_{\rm rot}
< 9/4$.
While our crude approximations are unable to definitively conclude that
such systems experience strong ZLK oscillations between the formation of the
first and second BH, it suffices as a first estimate of susceptibility.

\begin{figure}
    \centering
    \includegraphics[width=0.6\columnwidth]{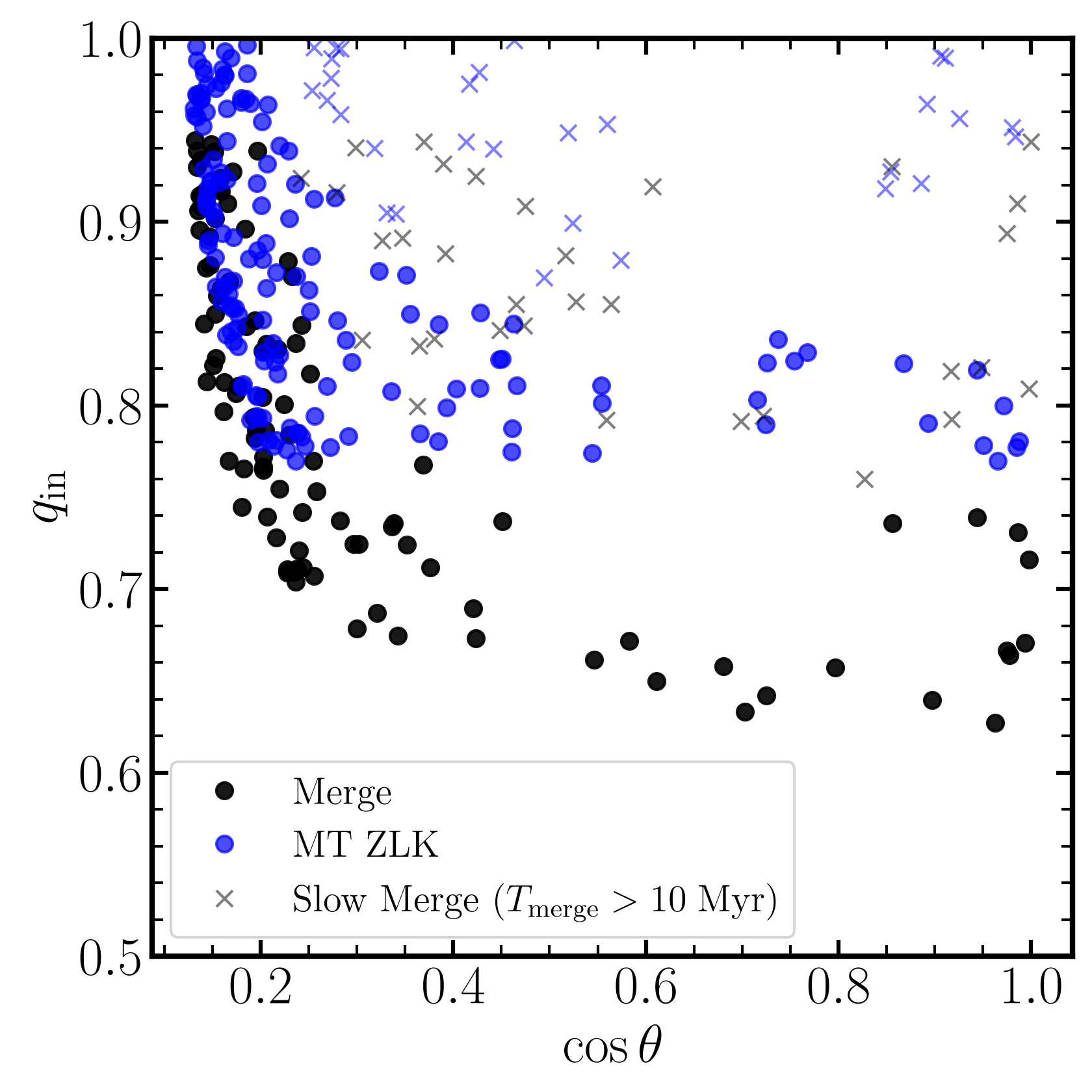}
    \caption{
    The properties of successfully merging BBH systems after the combined double
    mass transfer and orbital evolution summarized in
    Section~\ref{s:signatures}, corresponding to the MT prescription shown in
    the left panels of Fig.~\ref{fig:mt_res}.
    Black markers denote systems that successfully merge.
    Blue markers denote systems that merged but may be susceptible to ZLK during
    the double MT phase, during the interval between the formation of the first
    and second BHs (see Section~\ref{s:signatures}).
    Circular markers denote systems merging within $10\;\mathrm{Myr}$, while
    crossed markers denote systems merging between $10\;\mathrm{Myr}$ and
    $1\;\mathrm{Gyr}$, the
    maximum length of the orbital integration.
    }\label{fig:popsynth}
\end{figure}

The properties of the systems that successfully merge are shown in
Fig.~\ref{fig:popsynth}, where blue points denote systems that may experience
ZLK oscillations between the formation of the two compact objects based on the
discussion immediately above.
Such objects tend to have larger initial mass ratios, leading to more orbital
expansion during the formation of the first BH (Fig.~\ref{fig:mt_res}).
A clear anti-correlation between $q_{\rm in}$ and $\theta$ can be seen both in
the black points alone and in the entire population.
The inferred fraction of systems (among all misalignment angles $I$ between the
inner and outer orbits, uniformly weighted in $\cos I$) that successfully merge
within $1\;\mathrm{Gyr}$ is $\approx 6\%$, falling to $\approx 4\%$ when
restricting to systems merging within $10\;\mathrm{Myr}$; approximately $2/3$ of
these systems are susceptible to ZLK between the formation of the first and
second BHs.

\subsection*{Feasibility in Astrophysical Settings}

Above, we have shown that binary stars within the central regions of their SMBH
may form merging BBHs with correlations in their spin-orbit misalignment thanks
to their other physical properties due to two phases of stable MT\@.
In this section, we assess the feasibility of the proposed mechanism in typical
NSCs.

We first evaluate the dynamical lifetime of the binaries we consider.
In order to do so, we adopt an NSC stellar density profile in line with recent
studies \citep[e.g.,][]{rose2020_GC, jurado2024_kick}.
While the standard theoretical result, the Bahcall-Wolf profile
\citep{bahcallwolf_1976, binneytremaine_book} gives that $n(a_{\rm out}) \propto
a_{\rm out}^{-\alpha}$ with $\alpha = 7/4$, observations of the Milky Way
galactic center suggest a shallower
profile, $\alpha \sim 1.1$--$1.4$ \citep{gallegocano2018_nsc_alpha}, so we
instead adopt a value of $\alpha = 1.5$.
Then, when accounting for the standard $M$-$\sigma$ relation
\citep{tremaine2002}, we obtain the following stellar density profile
\begin{align}
    \rho (a_{\rm out})
        ={}& \frac{3 - \alpha}{2\pi}
            \frac{m_{\rm p}}{a_{\rm out}^3}
            \p{\frac{G\sqrt{m_{\rm p}M_0}}{\sigma_0^2 a_{\rm out}}}
                ^{\alpha - 3},\label{eq:nsc_rho}
\end{align}
where $M_0 = 10^8M_{\odot}$ and $\sigma_0 = 200\;\mathrm{km/s}$ are the
standard normalization, and $m_{\rm p}$ is the typical mass of stars in the
NSC\@.
In such a cluster, the velocity dispersion that our fiducial binary experiences
is
\begin{align}
    \sigma (a_{\rm out})
        &= \sqrt{\frac{Gm_3}{a_{\rm out}(1 + \alpha)}}\nonumber\\
        &= 700\;\mathrm{km/s}
            \p{\frac{m_3}{10^7M_{\odot}}}^{1/2}
            \p{\frac{a_{\rm out}}{0.04\;\mathrm{pc}}}^{-1/2}.
\end{align}
From this, we can calculate the characteristic binary dissociation/evaporation
timescale
\begin{align}
    t_{\rm evap}
        &= \frac{\sqrt{3}\sigma(a_{\rm out})}{
            32\sqrt{\pi}G\rho(r) a_{\rm in} \ln(\Lambda)}
            \frac{m_{12}}{m_{\rm p}}.\label{eq:tevap}
\end{align}
Here, $\ln \Lambda = 5$ is the Coulomb logarithm, and we take $m_{\rm p} \sim
m_{12} / 2$ to be the mass of flyby stars (corresponding to the simplifying
assumption that stars in the NSC all have mass $m_{\rm p}$ regardless of whether
they are in a binary).
For our fiducial parameters, this yields $t_{\rm evap} \simeq 10\;\mathrm{Myr}$.

Another constraint on the system properties arises from enforcing dynamical
stability of the triple.
For dynamical stability, we adopt the often used \citep{mardling2001tidal}
condition for dynamical stability:
\begin{align}
    \frac{a_{\rm out}}{a_{\rm in}}
        &\gtrsim
            2.8\p{1 + \frac{m_3}{m_{12}}}^{2/5}
            \frac{(1 + e_{\rm out})^{2/5}}{(1 - e_{\rm out})^{6/5}}
            \p{1 - 0.3\frac{I_{\rm tot, d}}{180^\circ}}.\label{eq:mardling}
\end{align}
For our fiducial parameters, the right-hand side of this expression is $\sim
1500$ (taking $I_{\rm tot, d} \simeq 90^\circ$, and adopting $m_{12} =
50M_{\odot}$, the final BH value), while the left-hand side is $\sim 4000$.
While alternative criteria are available (see \citealp{vynatheya2022_stab} for a
comparison of stability criteria as well as the most updated criteria), our
systems are sufficiently far from the dynamical stability boundary that the
detailed choice of criterion does not affect our result.

\begin{figure}
    \centering
    \includegraphics[width=\columnwidth]{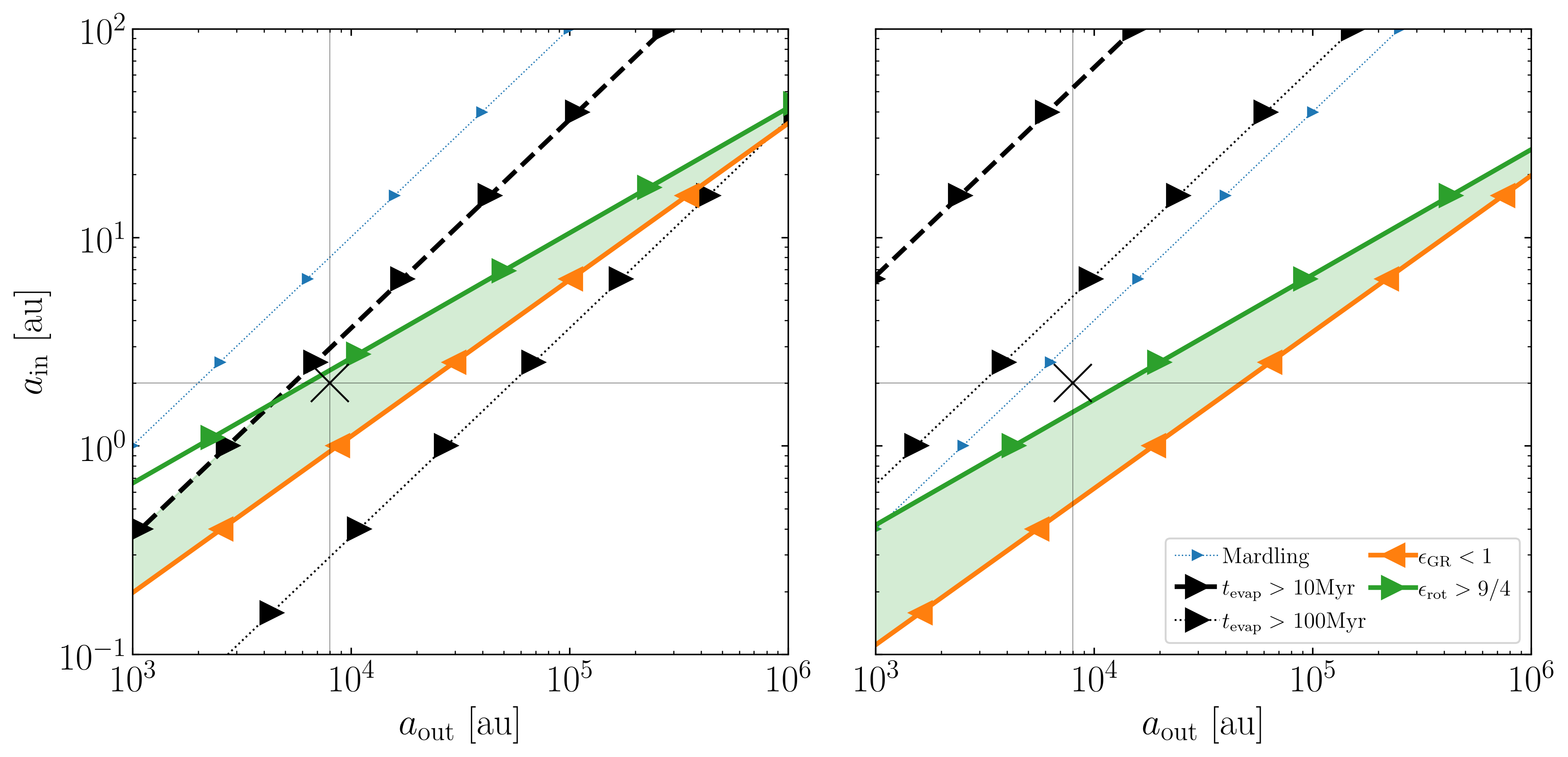}
    \caption{
    Parameter space available for the mechanism in this paper to form black hole
    binaries around a $10^7M_{\odot}$ SMBH (left) and around a $10^8M_{\odot}$
    SMBH (right).
    The blue dotted line denotes the condition for dynamical instability
    (Eq.~\ref{eq:mardling}), the black dashed and dotted lines denote the
    condition for binary dissociation/evaporation timescale to be longer than
    $10\;\mathrm{Myr}$ and $100\;\mathrm{Myr}$ (Eq.~\ref{eq:tevap}), and the
    green and orange lines denote the required conditions on the SRFs in the
    system (Eqs.~\ref{eq:epsrot_val} and~\ref{eq:eps_gr}).
    The green shaded region denotes the region of parameter
    space satisfying all of our constraints.
    For simplicity, we hold a few parameters fixed:     $m_{\rm 12, \star} =
    100M_{\odot}$, $m_{\rm 12} = 70M_{\odot}$, $q_{\rm in, \star} = q_{\rm in} =
    0.5$, $e_{\rm out} = 0.6$, and we take $R_\star = 10R_{\odot}$ and $k_{\rm
    q} = 0.02$, typical values for $\sim 60M_{\odot}$ stars.
    The fiducial parameters used in this study are labeled by the horizontal and
    vertical dotted lines, located at $a_{\rm in} = 2\;\mathrm{au}$ and $a_{\rm
    out} = 8000\;\mathrm{au} = 0.04\;\mathrm{pc}$.
    Note that, for the stellar density profiles given in Eq.~\eqref{eq:nsc_rho},
    there are $\sim 200$ stars in the green shaded region on the left and
    $\sim 40$ stars in the right.
    }\label{fig:timescales}
\end{figure}

Finally, for our mechanism, we require that $\epsilon_{\rm rot} > 9/4$ to
suppress eccentricity oscillations on the stellar phase, while we require that
$\epsilon_{\rm GR} \lesssim 1$ in order for eccentricity oscillations to
gradually induce merger of the BBH\@.
The combination of the four conditions listed above results in the parameter
space shown in Fig.~\ref{fig:timescales}.
While the parameter space satisfying all four requisite conditions is modest, it
is straightforward to dramatically expand the allowed parameter space by
invoking other mechanisms to suppress ZLK oscillations on the stellar binary
phase; we discuss this more below in Section~\ref{s:discussion}.
Note that for larger-mass SMBHs, the binary dissociation time is longer, but
this is due to a lower stellar density which has an adverse effect on event
rates.

Finally, we use the above results to briefly discuss the possible event rates
via this mechanism.
As the properties of binary mass transfer and the properties of NSCs (and
particularly binarity in NSCs) are not well constrained, we can do little more
than provide a very coarse estimate.
As a point of reference, we note that typical estimates of the number density of
NSCs are $\sim 10^{-2}\;\mathrm{Mpc}^{-3}$ in the $10^8M_{\odot}$ mass range
\citep{baldry2012_nscdensity, wright2017_nscdensity}.
By comparison to the latest BBH merger rate from GWTC-4.0
($14$--$26\;\mathrm{Gpc^{-3}yr^{-1}}$, \citealp{ligoo4a_updating}), we find that
we need $\sim 2\;\mathrm{merger\;NSC^{-1}\;Myr^{-1}}$.
Turning to our mechanism, the typical number of stars in the favorable region of
parameter space when taking typical stellar density profiles is $\sim 100$.
(Fig.~\ref{fig:timescales}), of which only some unknown fraction will have the
correct $a_{\rm in}$ to participate in the mechanism discussed here, and of
which only $\sim 10\%$ will have mutual inclinations with the outer tertiary
within $\sim [85^\circ, 95^\circ]$ (Fig.~\ref{fig:violin_bbhonly}).
Taken together, the proposed mechanism is likely a subdominant contribution
to the observed LVK mergers, though various efficiency-enhancing mechanisms
(e.g., outer binary reorientation, orbital evolution during stellar
binary evolution) can dramatically enhance this inferred rate.

\section{Discussion}\label{s:discussion}

Above, we have proposed a possible mechanism consisting of a combination of a
simplified binary stellar evolution and a subsequent dynamically-driven
merger mechanism that may play a role in explaining structures in the observed
parameters of merging binary black holes.
In this section, we discuss these results in the context of the existing
literature.
We first point out the unique constraints our mechanism aims to satisfy, then we
discuss the numerous points of further study that would strengthen the various
assumptions made throughout the current work.

\subsection{Comparison to Other Mechanisms}

We begin by reviewing the most recent results on the spin statistics as of
GWTC-5.0 \citep{ligo_gwtc5intro}, which are not significantly different to those
reported as of GWTC-4.0 \citep{ligoo4a}.
First, it has now become clear that BH spins are preferentially more aligned
than anti-aligned with their orbits \citep{ligo_o3b_props,
callister2024_gwreview, ligoo4a_props, ligo_gwtc5pop}.
However, the detailed shape of the $\theta$ distribution continues to be
refined:
statistical analyses uncover a tentative, broad peak in the $\cos \theta$
distribution around $\cos\theta \in [-0.3, 0.9]$
\citep{vitale2022_spinasyoulike, edelman2023_spin, callister2024_paramfree,
baibhav2024_spin, ligoo4a_props} that may be more pronounced in the low-mass
bulk population of BBH mergers \citep{stegmann2026_triples} and a general lack
of significantly anti-aligned systems ($\cos\theta \lesssim -0.7$;
\citealp{ligoo4a_props}).
Both of these features are consistent with our mechanism, which preferentially
produces systems with $\cos\theta > 0$ (see e.g.,
Fig.~\ref{fig:violin_bbhonly}), though mildly negative $\cos\theta$ can easily
be produced as well \citep{LL18, su2021_lk90}.

At the same time, the nature of the $q$-$\chi_{\rm eff}$ anti-correlation
\citep{callister2021_qchi} has changed character over time: while there is
growing evidence for correlation between the two quantities
\citep{ligo_o3b_props, ligoo4a_props}, an emerging debate is whether this
correlation is with the mean or width of the $\chi_{\rm eff}$ distribution
\citep{heinzelbisco2024_60peak, ligoo4a_props}.
This work provides a plausible dynamical origin for an anticorrelation between
$q$ and the mean $\chi_{\rm eff}$ (see Fig.~\ref{fig:popsynth}).
A correlation between $q$ and the width of the $\chi_{\rm eff}$ distribution can
also be obtained dynamically \citep[e.g.,][]{LL18}.
Note that, as of GWTC-5.0, the data seem to prefer a anti-correlation of $q$
with the width of the $\chi_{\rm eff}$ distribution rather than its mean
\citep[see Appendix C of][]{ligo_gwtc5pop}.

While the efficiency of the mechanism as presented is modest
(Section~\ref{s:signatures}), it also poses several advantages compared to existing
mechanisms in the literature:
\begin{itemize}
    \item Compared to previous studies of tertiary-induced mergers, our
        mechanism has a physical mechanism for the BBH's initial spin-orbit
        alignment, namely the preceding MT phase.
        This primordial alignment is put in by fiat in existing studies of
        tertiary-driven mergers that report strongly-peaked spin distributions
        \citep{antonini2018precessional, LL18, su2021_lk90}, and the spin-orbit
        misalignments become isotropic when these initial conditions are not
        well-obeyed \citep{fragione2020, yu2020_spin}.

        Naively, it would be expected that the orbital orientations of wider
        stellar binaries ($a_{\rm in} \gtrsim 100\;\mathrm{au}$) likely have no
        correlation with the stellar spin directions due to the vast difference
        in spatial scales.
        However, recent observations of exoplanetary systems suggest that the
        spins of stars may sometimes be aligned with the orbital planes of their
        binary companions out to several hundreds of AU, likely as a consequence
        of their formation \citep{rice2024spinorbitorbit}.
        As such, we can speculate that the assumption of initial spin alignment
        in previous studies may be somewhat justified.

        Our work also differs from studies where the tertiary is a stellar-mass
        companion.
        In such systems, the octupole-order effects in the hierarchical
        expansion (Eq.~\ref{eq:phi_oct}) are much stronger and more efficiently
        produce extreme eccentricities in BBHs with unequal mass ratios
        \citep{ford2000secular, naoz2016eccentric, silsbee2017lidov}.
        Some studies find that such an enhancement is incompatible with LVK data
        \citep{su2021_massratio}, though comparison with other studies suggests
        that this conclusion may be dependent on the inner binary separation
        \citep{martinez2021mass}.
        When the tertiary is an SMBH, the octupole-order effects are much
        weaker due to the large value of $a_{\rm out}$ (Eq.~\ref{eq:eps_oct}).

        Finally, a common obstacle in tertiary-induced mergers is modification
        (and even unbinding) of the inner and outer orbits during BH formation,
        both due to natal kicks and Blaauw kicks \citep[e.g.,][]{fragione2020,
        liu2021hierarchical, su2024_superthermal}.
        In our mechanism, the large Keplerian orbital velocities of both the
        inner and outer orbits ($v_{\rm in} \sim 200\;\mathrm{km/s}$ and $v_{\rm
        out} \sim 10^3\;\mathrm{km/s}$) allow us to neglect the effect of natal
        kicks during the formation of the BHs (which are poorly constrained but
        are typically assumed to be $\lesssim 100\;\mathrm{km/s}$ if not
        substantially lower, e.g.\ \citealp{rodriguez2016binary}).

    \item Compared to formation channels where BBHs are dynamically assembled,
        our channel provides mechanisms for introducing preferred spin
        orientations (via MT) and matching the correlations found in the LVK
        observations.
        The spins of BHs in dynamically assembled binaries are typically thought
        to be isotropically oriented at formation \citep{costa2023_review}, leading
        to a uniform distribution of $\cos\theta$ (and a triangular distribution
        in $\chi_{\rm eff}$, \citealp{fragione2020}).
        Note that late-stage general relativistic effects can drive the in-plane
        components of BH spins from isotropy towards [anti-]alignment
        \citep{schnittman2004, gerosa2013, gerosa2023_precessioncode}, but this
        does not affect $\chi_{\rm eff}$.
        However, recent work suggests that BBHs in dense stellar environments
        may have their component spins realigned with their orbits if they
        experience collisions with nearby stars \citep{kiroglu2025_gcspin}.
        Furthermore, a natural correlation between $q$ and the width of the
        $\chi_{\rm eff}$ distribution can naturally be obtained via hierarchical
        mergers in such dense environments \citep{vijaykumar2026_massratio}.

    \item Compared to BBH formation via isolated binary evolution, our mechanism
        does not require rather precise amounts of orbital shrinkage via a common
        envelope phase.
        The stellar binaries we consider experience only moderate orbital
        evolution during their two phases of stable MT\@.
        On the other hand, recent works suggest that the LVK spin signatures,
        including the $q$-$\chi_{\rm eff}$ correlation, may be reproducible with
        isolated binary evolution alone \citep{olejak2024_unequalmass,
        banerjee2024_olejak_qchi, baibhav2024_spin}.

    \item Compared to BBH formation and merging in the disks of active galactic
        nuclei (AGN), our mechanism is less sensitive to the details of the
        local environment and highly uncertain hydrodynamical effects
        \citep[see][for a recent review]{lai2023_circumbinaryreview}.
        Nevertheless, the AGN channel currently provides many prospects for
        reproducing the observed spin signatures as well as the high-mass end of
        merging BBH systems, and its detailed quantitative predictions are
        still being better understood \citep[e.g.,][]{li2021, wang2021_agn,
        mckernan2022_qchi, santini2023_disk, cook2024_mcfacts2}.
        Moreover, if indeed SMBHs grow due to episodic bursts of accretion from
        AGN disks that persist only for a characteristic lifetime $\sim
        0.1\;\mathrm{Myr}$ \citep[as suggested by][]{schawinsky2015_agnflicker},
        then the gas-free dynamics explored in this work and the AGN disk-driven
        mechanism may be complementary, alternately active in driving BBH
        mergers in the innermost regions of galaxies.
\end{itemize}

Finally, it must be noted that the entire LVK catalog may arise from multiple
formation channels, and recent work presents growing evidence for this
possibility \citep{zevin2021, kimball2021, li2024_2channel, li2024_2channelb,
hussain2024_2pop, li2025_2channel, banagiri2025_threepop, farah2026_subpop,
ray2026_threepop, plunkett2026_subpop}.
For instance, recent results using the data from GWTC-5.0 suggest that the
structure in the $q$--$\chi_{\rm eff}$ distribution may be due to a combination
of a slowly-spinning bulk centered on $\chi_{\rm eff} \approx 0$ and a
positive-skewed $\chi_{\rm eff}$ distribution at specific primary masses
originating from a wide range of merger channels
\citep{alvarezlopez2026_gwtc5bulkzero}.

\subsection{Caveats and Future Work}\label{ss:disc}

Due to the broad scope of our work and the approximate treatments contained
within it, there are numerous caveats and avenues for future work.
We first briefly enumerate them as follows for convenience:
(i) In deriving the relation between the initial and final spin orientations of
tertiary-induced mergers, we have assumed initial spin-orbit alignment of the
BBH upon its formation, and we have assumed that the tertiary is sufficiently
distant that single-averaged and octupole-order effects can be neglected,
allowing for the application of Eq.~\eqref{eq:naive}.
(ii) In our subsequent results, we have shown that a population of stellar
binaries with fixed initial $a_{\rm in, \star}$ and varying $q_{\rm in, \star}$
undergoing two phases of stable mass transfer can give rise to correlated values
of $q_{\rm in}$ and $a_{\rm in}$ for the subsequent BBH formed at the end of
binary stellar evolution when assuming a simple mass transfer prescription that
does not vary for different stellar masses and separations.
(iii) In studying the possibility of nuclear star clusters as natural
astrophysical locations for this process to occur, we suggest that potential ZLK
oscillations during the stellar phase are suppressed by apsidal precession due
to the stars' rotational bulges and assume that the stars in nuclear star
clusters are born in situ and all in binaries.
We discuss the motivations and consequences of these assumptions in greater
detail below.

First, we discuss the details of the spin and orbital evolution.
In expanding the interaction potential to octupole order and performing a double
averaging over the inner and outer orbits, we have neglected terms of the
hexadecapole order and higher \citep{will2017_hexa, will2021_quad2,
conway2024_32} as well as neglected effects due to the breakdown of the outer
orbit's averaging \citep{luo2016, grishin2018SA}.
For the extreme orbital hierarchies we consider ($a_{\rm in} / a_{\rm out}
\sim P_{\rm in} / P_{\rm out} \lesssim 10^{-3}$), both of these approximations
are well-justified: indeed, the evolution is dominated by the quadrupole-order
ZLK evolution, and both the octupole-order and Brown's Hamiltonians already
only contribute negligibly.
However, if the spin dynamics considered in this work and preceding ones
\citep{LL18, su2021_lk90} is to be extended to lower-mass tertiary companions or
to other regions of the nuclear star cluster where these effects are not so
negligible, the spin adiabatic invariant must be quantitatively re-examined, as
the ZLK cycles are no longer sufficiently regular to justify the assumptions of
Section~\ref{ss:ad_invar}.
A similar issue may likely arise due to the breakdown of double averaging alone,
a phenomenon coined \emph{relativistic phase space diffusion}
\citep{hamilton2024_rpsd1} that can also lead to significant variations to ZLK
cycles \citep{klein2026_rpsd2}.
Generally, as these higher-order effects become marginally important, one
expects conservation of the adiabatic invariant $\theta_{\rm eff}$
(Eq.~\ref{eq:naive}) to become poorer, broadening the resulting $\theta$
distribution without significant effects on the mean \citep[e.g. Fig.~20
of][]{LL18}.

Next, we discuss our treatment of binary stellar evolution (BSE).
Note that, notwithstanding the many simplifying assumptions we have made, the
key objective with our evolution was to determine the strength of any
correlation between the BBH's initial semi-major axis and mass
ratio\footnote{
Note that a positive correlation between the binary semi-major axis and mass
ratio has also been reported in common envelope binary evolution
\citep{sand2020_correlation, borges2026_cecorrelation}.
This suggests that stable mass transfer may not be a necessary component for
reproducing the final correlations in this work.
}.
We encourage investigation of such correlations with more sophisticated BSE
models, in particular with modern binary evolution codes such as
\texttt{POSYDON} \citep{posydon_code, posydon2_code} that correctly capture the
double mass transfer we invoke (as opposed to older codes based on
\citealp{hurley2002} that likely overpredict the onset of common envelope
evolution, e.g.\ \citealp{gallegosgarcia2021_2mt}).
Even so, we discuss some uncertainties in our parameter choices below.
First, we have chosen a moderate amount of wind-powered mass loss ($\eta_{\rm
wind} = 0.2$) reflecting the recent prevailing wisdom that winds are likely
subdominant to binary mass transfer in stripping massive stars
\citep{smith2014_masslossrev, vinciguerra2020_MT}.
On the other hand, $\eta_{\rm wind}$ is likely larger for higher-metallicity
stars, and the mean metallicity in the Milky Way's NSC appears to be solar to
super-solar \citep{lepine2011metal, do2015_mwmetal, schodel2020_sstarmetal}, so
we have chosen not to fully neglect it.
The details of whether MT in a binary is stable or unstable are uncertain, and
likely depend on both mass ratio and orbital separation
\citep[e.g.,][]{schneider2015_mt, schurmann2024_smt} as well
as the specifics of stellar evolution \citep[e.g.,][]{passy2012_giantmassloss,
pavlovskii2015, agrawal2020_msrad, klencky2022_giantMT, temmink2023_MTstab}.
Furthermore, note that while our adopted accretion efficiency during the first
stable MT phase ($\epsilon_2 = 0.5$) is standard \citep{meurs1989,
dominik2012double, vanson2022_analmt, schurmann2024_smt}, the most physical
values for $\epsilon_2$ may be substantially smaller if rotationally-enhanced
stellar winds efficiently re-emit material from the accretor during the MT phase
(\citealp{MESA_3, posydon2_code}, but see \citealp{vinciguerra2020_MT}).
Additionally, $\epsilon_2$ may vary as a function of mass ratio
\citep[e.g.,][]{schneider2015_mt},
and more sophisticated population synthesis codes contain more extensive and
well-calibrated, albeit less interpretable, results for the effect of double
mass transfer \citep[e.g.,][]{cosmic_code, posydon2_code}.
Nevertheless, Fig.~\ref{fig:mt_res} suggests that the correlation we seek may be
relatively insensitive to these uncertainties considered above.
As another caveat to our treatment of the stellar binary phase, we remark that
our justification of suppressed ZLK oscillations on the stellar phase relies on
evaluation at ZAMS (see Section~\ref{ss:bse_srf}).
Of course, these parameters change as the stars evolve: stellar rotation rates
slow down as they expand, but their radii also grow, having opposite effects on
the suppression effect (Eq.~\ref{eq:epsrot_val}).
In general, one might expect the net effect of post-MS expansion to be an
increase suppression of ZLK as the star expands, since its rotational quadrupole
is concentrated at its surface \citep{eggleton2006book}.
Finally, our neglect of BH natal kicks was made under the assumption that the
kick velocities are $\lesssim 100\;\mathrm{km/s}$, in agreement with both
standard population synthesis prescriptions \citep{rodriguez2016binary,
Giacobbo2020} and recent observational constraints \citep{kimball2023kick,
vignagomez2023natal, banagiri2023, burdge2024v404}.

One assumption made in this work, for simplicity, is a unimodal initial stellar
semi-major axis distribution $a_{\rm in, \star}$.
While certainly oversimplified, a preferred semi-major axis for BBH progenitors
may be well-justified: compact binaries experience strong SRFs that suppress ZLK
oscillations, while wide binaries can merge on the main sequence after reaching
sufficiently high eccentricities \citep[e.g.,][]{stephan2016_gcmergers,
leigh2016_mergers, fragione2019_bluestragglers}.
Furthermore, star formation is thought not to be particularly efficient at
$\lesssim \mathrm{au}$ scales \citep{offner2023review}, introducing another
characteristic scale to the $a_{\rm in, \star}$ distribution.
Nevertheless, comparative studies with different $a_{\rm in, \star}$
distributions should certainly be incorporated before quantitative comparisons
to observation are made, which we defer to future work.

Finally, we discuss the uncertainties regarding the feasibility of our mechanism
in galactic center environments.
While the formation of NSCs remains an open question (being primarily attributed
to either infalling globular clusters or in situ formation, see
\citealp{neumayer2020} for a recent review), the young inferred age of stars in
the Milky Way NSC \citep{figer2004_younggc, rossa2006_younggc2, do2015_mwmetal}
and several observed binaries near Sgr A$^*$ \citep[with parameters similar to
our fiducial ones]{martins2006_sga1, pfuhl2014_sga2, peissker2024_binarySgr}
suggest that there may be a continually-replenished population of massive
stellar binaries like those considered in this work \citep[but see
also][]{chu2023_sgrabin}.

In addition, we remark that line-of-sight acceleration of the BBH center-of-mass
introduces corrections to the GW waveform \citep{yunes2011_smbhlos,
bonvin2017_smbhlos}, an important and testable prediction of BBH merger
mechanisms in the vicinity of SMBHs \citep{inayoshi2017_comaccel} that also
carries implications for comparison to the inferred properties of the merging
BBHs, which use isolated binary waveforms \citep{ligo_gwtc5intro}.
Here, we report the effect of these results on our mechanism using the results
of \citet{vijaykumar2023_waltz}, who perform detectability studies of
the center-of-mass acceleration experienced by the BBH in units of $\dot{v}_{\rm
com}/c
[\mathrm{s}^{-1}]$.
For our fiducial parameters, the binaries we study experience an acceleration
$\dot{v}_{\rm com} / c \sim 3 \times 10^{-12}\p{m_3 / 10^7 M_{\odot}} \p{a_{\rm
out} / 0.04\;\mathrm{pc}}^{-2}\;\mathrm{s}^{-1}$.
Such a small acceleration is suggested to remain undetectable even to the next
generation of GW detectors (Cosmic Explorer and Einstein Telescope,
specifically) \citep{vijaykumar2023_waltz}.
While we have also considered broader regions of parameter space (see
Fig.~\ref{fig:timescales}), regions corresponding to significantly larger
$\dot{v}_{\rm com} /
c$ generally will result in either mergers on the stellar phase (due to
$\epsilon_{\rm rot} < 9/4$) or suffer from a lack of available binary stars (the
stellar number density given in Eq.~\ref{eq:nsc_rho}), often both.
On the other hand, the acceleration parameter \citep{inayoshi2017_comaccel} for
our fiducial parameters takes on value $\epsilon \equiv 3000\p{m_3 / 10^7
M_{\odot}} \p{a_{\rm out} / 0.04\;\mathrm{pc}}^{-2}$, which appears to roughly
marginally detectable for systems with chirp masses $\sim 40M_{\odot}$ detected
by the combined LISA+LIGO pipeline \citep{inayoshi2017_comaccel}.
For completeness, we remark that gravitational redshift
\citep{chen2019_gravredshift} is not expected to be important given that
$0.04\;\mathrm{pc} \sim 10^{5}R_{\rm Sch,7}$ where $R_{\rm Sch,7}$ is the
Schwarzschild radius of a $10^7M_{\odot}$ SMBH\@.

It is worth noting that, in the dense stellar environments of NSCs, a key
constraint of the mechanism as proposed in this work can be relaxed.
The small parameter space shown in Fig.~\ref{fig:timescales} is most strongly
constrained by the requirement that the primordial stellar binary avoid ZLK via
strong SRFs, while the resulting BBH merges efficiently via ZLK-driven
eccentricity oscillation.
We made this choice in order to most clearly illustrate the properties of the BH
spin distributions that can result from our mechanism.
However, it can instead be the case that the initial stellar binary is either
initially compact or aligned with the outer orbit, and relaxation effects drive
the binary into a ZLK-active configuration.
This can be either due to two-body relaxation that softens the binary orbit
\citep[e.g.,][]{collins2008_levyflight, naoz2022_relax, wintergranic2024,
hamiltonmodak2024_closeencounters} or due to resonant relaxation
\citep{rauch1996_vrr, hamers2018_vrr}.
We defer the inclusion of these effects to future work, as the impulsive nature
of each successive close encounter likely complicates the spin evolution of
the BBH components, broadening their observed distributions but potentially
retaining some fundamental correlation driven by the mechanism introduced here.

\subsection{Conclusion}\label{ss:summary}

In this paper, we have studied the formation of merging binary black holes
(BBHs) formed from stellar binaries in the vicinity of a supermassive black
hole.
We have shown that the combination of stellar and coupled spin-orbit evolution
can give rise to unexpected correlations in the observed spin distributions of
BBHs.
To be precise, the analytically tractable spin evolution of a BBH from formation
to merger (Section~\ref{s:spindyn}, and Eq.~\ref{eq:naive}) combines with the
stellar evolution of the progenitor binary (Section~\ref{s:stellar_evol}) to
yield an anti-correlation between the mass ratio $q$ and the cosines of the
spin-orbit misalignment angles $\cos \theta_{1,2}$
 of the merging BBHs
(Section~\ref{s:signatures} and Fig.~\ref{fig:popsynth}).
This correlation is reminiscent of the observed $q$-$\chi_{\rm eff}$
correlation \citep{callister2021_qchi,
callister2024_gwreview}\footnote{
Direct comparison of the two trends is complicated by the unknown BH
spin magnitudes ($\chi_{1, 2}$ in Eq.~\ref{eq:chieff}).
These are expected to be slow based on theoretical arguments
\citep{fuller2019most, marchant2024_spins}, but other studies suggest more rapid
rotation based on helioseismic constraints \citep{eggenberger2019_rotcouple},
and inferences from the LVK data find $\chi \sim 0.2$ \citep{ligoo4a_props}.}.
Altogether, our work represents an important step in identifying possible
sources of structure in the distribution of BH spins formed in
gravitationally-driven BBH merger channels.


\vspace{6pt}

\funding{
Y.S.\ acknowledges support by the Lyman Spitzer Jr.\ Postdoctoral Fellowship at
Princeton University and by the Natural Sciences and Engineering Research
Council of Canada (NSERC) [funding reference CITA 490888-16].
This work used NSF ACCESS allocations on Expanse at the San Diego Supercomputer
Center (SDSC) at UC San Diego through allocation PHY230201 from the Advanced
Cyberinfrastructure Coordination Ecosystem: Services \& Support (ACCESS) program
\citep{accesscitation_2023}, which is supported by U.S. National Science
Foundation grants \#2138259, \#2138286, \#2138307, \#2137603, and \#2138296.
}

\dataavailability{
The raw data supporting the conclusions of this article will be made available
by the author on request.
}

\acknowledgments{
We thank the reviewers for helpful and insightful comments that improved the
quality of this paper.
YS thanks Lieke van Son for many long discussions about the intricacies of
binary evolution. YS also thanks Mark Dodici, Will Farr, Saavik Ford, Evgeny
Grishin, Chris Hamilton, Vicky Kalogera, Barry McKernan, Dong Lai, Bin Liu,
Shaunak Modak, Carl Rodriguez, Connar Rowan, Mor Rozner, Eliot Quataert, Lucas
M.\ de S\'a, and Sylvia Biscoveanu for helpful comments.
}

\conflictsofinterest{
The authors declare no conflicts of interest.
}





\appendixtitles{Yes} 
\appendixstart
\appendix
\section{Equations of Motion}\label{app:eom_zlk}

Here, we give the equations of motion describing the double-averaged orbital
evolution of the black hole triple system to octupolar order
\citep[e.g.,][]{naoz2016eccentric}, including
``Brown's Hamiltonian'' \citep{tremaine2023brown, grishin2024_brown}, general
relativistic apsidal precession of the inner binary \citep{liu2015suppression},
and gravitational wave (GW) emission \citep{peters1964}.
We adopt the common vectorial formulation, where the inner and outer orbits are
described by their orbital angular momentum and eccentricity vectors
\citep[e.g.,][]{LL18, su2021_lk90}
\begin{align}
    \bm{L}_{\rm in}
        &= L_{\rm in}\bm{j}_{\rm in}\nonumber\\
        &= \mu_{\rm in} \sqrt{Gm_{12}a_{\rm in}}
            \;\bm{j}_{\rm in},\\
    \bm{e}_{\rm in} &= e_{\rm in}\uv{e}_{\rm in},\\
    \bm{L}_{\rm out}
        &= \mu_{\rm out} \sqrt{Gm_{123}a_{\rm out}}
            \;\bm{j}_{\rm out},\\
    \bm{e}_{\rm out} &= e_{\rm out}\uv{e}_{\rm out}.
\end{align}
Here, $\mu_{\rm out} \equiv m_{12}m_{3} / m_{123}$ and $m_{123} = m_{12} + m_3$,
and $\bm{j}_{\rm in} \equiv j_{\rm in}\jhat_{\rm in}$ where $j_{\rm in}^2 \equiv
1 - e_{\rm in}^2$ (and analogously for $j_{\rm out}$).

Then, the interaction potential between the two orbits is given by \citep{LL18,
tremaine2023brown, grishin2024_brown}
\begin{align}
    \Phi ={}& \Phi_{\rm quad} + \Phi_{\rm oct} + \Phi_{\rm B},\\
    \Phi_{\rm quad}
        ={}& \Phi_0 \Big[
            1 - 6e_{\rm in}^2
                - 3(\bm{j}_{\rm in} \cdot \jhat_{\rm out})^2
                + 15(\bm{e}_{\rm in} \cdot \jhat_{\rm out})^2
                \Big],\\
    \Phi_{\rm oct}
        ={}&
            \frac{15}{8}\epsilon_{\rm oct}\Phi_0\Big[
                \p{\bm{e}_{\rm in} \cdot \uv{e}_{\rm out}}\nonumber\\
                &\times \p{
                    8e_{\rm in}^2
                    -1
                    + 5\p{\bm{j}_{\rm in} \cdot \jhat_{\rm out}}^2
                    - 35\p{\bm{e}_{\rm in} \cdot \jhat_{\rm out}}^2
                }
                \nonumber\\
            &+
                10
                \p{\bm{e}_{\rm in} \cdot \jhat_{\rm out}}
                \p{\bm{j}_{\rm in} \cdot \uv{e}_{\rm out}}
                \p{\bm{j}_{\rm in} \cdot \jhat_{\rm out}}
        \Big],\label{eq:phi_oct}\\
    \Phi_{\rm B}
        ={}&
            \frac{3(3 + 2e_{\rm out}^2)}{8}\epsilon_{\rm SA} \Phi_0
                \p{\bm{j}_{\rm in} \cdot \jhat_{\rm out}}\nonumber\\
                &\times \p{
                    24 e_{\rm in}^2
                    - 15 \p{\bm{e}_{\rm in} \cdot \jhat_{\rm out}}^2
                    - \p{\bm{j}_{\rm in} \cdot \jhat_{\rm out}}^2
                    + 1
                }.
\end{align}
We have defined the standard quantities \citep[e.g.,][]{liu2015suppression, grishin2024_brown}
\begin{align}
    \Phi_0
        &= \frac{Gm_1m_2m_3a_{\rm in}^2}{8m_{12}a_{\rm out}^3j_{\rm
        out}^{3}},\\
    \epsilon_{\rm oct}
        &\equiv \frac{m_1 - m_2}{m_{12}}
            \frac{a_{\rm in}}{a_{\rm out}}
            \frac{e_{\rm out}}{1 - e_{\rm out}^2},\label{eq:eps_oct}\\
    \epsilon_{\rm SA}
        &\equiv
            \p{\frac{m_3^2}{m_{12}m_{123}}}^{1/2}
            \p{\frac{a_{\rm in}}{a_{\rm out}j_{\rm out}^2}}^{3/2}.
\end{align}
Finally, $\Phi_{\rm GR}$ describes general relativistic pericenter advance, and
is given by
\begin{equation}
    \Phi_{\rm GR}
        =
            \frac{3Gm_{12}}{c^2a_{\rm in}j_{\rm in}} n_{\rm in}
\end{equation}
where $n_{\rm in} = \sqrt{Gm_{12} / a_{\rm in}^3}$ is the mean motion of the
inner binary.
As described in the main text, $\Phi$ can be used to generate the equations of
motion for the conservative dynamics via the Milankovich equations
(Eq.~\ref{eq:milankovich_L}--\ref{eq:milankovich_e}), most robustly via computer
algebra.

Finally, the equations of motion describing GW emission are well-known
\citep{peters1964}
\begin{align}
    \at{\rd{\bm{L}_{\rm in}}{t}}_{\rm GW}
        &= -\frac{32}{5}
            \frac{G^{3}\mu_{\rm in} m_{12}^{2}}{c^5a_{\rm in}^{4}j_{\rm in}^5}
            \p{1 + \frac{7e_{\rm in}^2}{8}}\bm{L}_{\rm in},\label{eq:dadt_gw}\\
    \at{\rd{\bm{e}_{\rm in}}{t}}_{\rm GW}
        &=
            -\frac{304}{15}
                \frac{G^3\mu_{\rm in}m_{12}^2}{
                c^5a_{\rm in}^4 j_{\rm in}^5}
                \p{1 + \frac{121e_{\rm in}^2}{304}}
                \bm{e}_{\rm in}.
\end{align}
Since GW emission is not conservative, its effect must be added by hand to the
result of the Milankovich equations.

\section{Derivation of Spin Evolution Adiabatic
Invariant}\label{app:spin_adinvar}

In this section, we present a simplified qualitative derivation illustrating the
evolution of the spin-orbit misalignment angle $\theta$ during the course of a
tertiary-driven merger; see \citet{su2021_lk90} for a more rigorous discussion,
and see also \citet{LL18, yu2020_spin} for other discussions of this mechanism.

For simplicity, we will assume that $\bm{L}_{\rm out}$ is approximately fixed
and that $\epsilon_{\rm oct}$, $\epsilon_{\rm SA} \ll 1$, well satisfied for our
fiducial parameters.
Note that, as $\uv{s}$ precesses about $\jhat_{\rm in}$ due to de Sitter
precession, the orientation of $\jhat_{\rm in}$ itself also changes over
the characteristic quadrupole-order ZLK timescale
\begin{align}
    \frac{1}{t_{\rm ZLK}}
        ={}& n_{\rm in}\frac{m_3}{m_{12}}
            \p{\frac{a_{\rm in}}{\tilde{a}_{\rm out}}}^3,\\
        ={}& \frac{1}{10^4\;\mathrm{yr}}
            \p{\frac{m_3}{10^7M_{\odot}}}
            \p{\frac{m_{12}}{70M_{\odot}}}^{-1/2}\nonumber\\
        &\times \p{\frac{a_{\rm in}}{2\;\mathrm{au}}}^{3/2}
            \p{\frac{\tilde{a}_{\rm out}}{6000\;\mathrm{au}}}^{3/2}.
\end{align}
where $\tilde{a}_{\rm out} \equiv a_{\rm out}j_{\rm out}$.
Specifically $\jhat_{\rm in}$ both precesses and \emph{nutates} (changing $I$)
about $\jhat_{\rm out}$.

Depending on the value of $\Omega_{\rm dS}t_{\rm ZLK}$, there can be two regimes
of evolution:
\begin{itemize}
    \item First, if $\Omega_{\rm dS}t_{\rm ZLK} \ll 1$ (GR effects are weak),
        then de Sitter precession is too slow to drive precession of $\uv{s}$
        about the instantaneous orientation of the rapidly-varying $\jhat_{\rm
        in}$.
        As a result, $\uv{s}$ instead precesses about some suitably
        time-averaged axis, which we discuss below.

    \item Second, if $\Omega_{\rm dS} t_{\rm ZLK} \gg 1$ (GR effects are
        strong), then de Sitter precession is sufficiently rapid that $\uv{s}$
        can efficiently follow the slow variations of $\jhat_{\rm in}$.
        As a result, $\theta$ remains approximately constant, as an action of
        the spin Hamiltonian \citep{landau_lifshitz}.
\end{itemize}
In the first regime, the appropriate time averaging can be identified by
treating the spin evolution as an iterative map over successive ZLK cycles.
\begin{align}
    \frac{\uv{s}_{k + 1} - \uv{s}_k}{P_{\rm ZLK}}
        &=
            \int\limits_{t_0}^{t_0 + P_{\rm ZLK}}
                \Omega_{\rm dS}\jhat_{\rm in}
                \times \uv{s}\;\mathrm{d}t\nonumber\\
        &\approx
            \s{\int\limits_{t_0}^{t_0 + P_{\rm ZLK}}
                \Omega_{\rm dS}\jhat_{\rm in}\;\mathrm{d}t}
                \times \uv{s}_k\nonumber\\
        &\equiv
            \ev{\Omega_{\rm dS}\jhat_{\rm in}} \times \uv{s}_k,
\end{align}
where the angle brackets denote averaging over a ZLK cycle.
Thus, over timescales $\gg P_{\rm ZLK}$, the average spin evolution can be
described by
\begin{align}
    \ev{\rd{\uv{s}}{t}}
        &=
            \ev{\Omega_{\rm dS}\jhat_{\rm in}} \times \uv{s}.
            \label{eq:ev_dsdt_in}
\end{align}
However, this expression is still difficult to analyze, since the orientation of
$\ev{\,\jhat_{\rm in}}$ still varies on timescales $\sim P_{\rm ZLK}$:
Namely, while its nutation has been eliminated by the averaging, it will
continue to precess about $\jhat_{\rm out}$ at a rate dominated by the
quadrupolar-order ZLK evolution \citep[e.g.,][]{liu2015suppression, su2021_lk90}:
\begin{align}
    \rd{\ev{\,\jhat_{\rm in}}}{t}
        &\approx
            \ev{\Omega_{\rm ZLK}}
            \p{\,\jhat_{\rm out} \times \ev{\,\jhat_{\rm in}}}
            ,\\
    \Omega_{\rm ZLK}
        &\equiv
            \frac{3}{4t_{\rm ZLK}}
            \frac{\cos I (5e_{\rm in}^2\cos^2\omega_{\rm in} - 4e_{\rm in}^2 - 1)
                }{j_{\rm in}}\label{eq:def_Omega_ZLK}.
\end{align}
where $\omega_{\rm in}$ is the argument of pericenter of the inner binary.
To eliminate this precession, we perform a change of reference frame from
Eq.~\eqref{eq:ev_dsdt_in} to the co-precessing frame, obtaining
\begin{align}
    \ev{\rd{\uv{s}}{t}}_{\rm co-pre}
        &=
            \s{\ev{\Omega_{\rm dS}\jhat_{\rm in}}
               - \ev{\Omega_{\rm ZLK}}\jhat_{\rm out}} \times \uv{s}\nonumber\\
        &\equiv
            \bm{\Omega}_{\rm eff} \times \uv{s} \label{eq:dsdt_Weff}.
\end{align}
In this reference frame, $\bm{\Omega}_{\rm eff}$ varies little over successive
ZLK cycles, and only evolves gradually due to GW emission\footnote{Note that
large $\epsilon_{\rm oct}$, large $\epsilon_{\rm SA}$, or spin-orbit resonances
also result in substantial variation of $\bm{\Omega}_{\rm eff}$ between
successive ZLK cycles. These all result in non-conservation of $\theta_{\rm
eff}$, which leads to unpredictable spin evolution \citep{LL18, LL19,
su2021_lk90}.}.
Thus, as long as the GW-driven evolution of the system (Eq.~\ref{eq:dadt_gw}) is
much slower than $P_{\rm ZLK}$, the angle
\begin{equation}
    \theta_{\rm eff}
        \equiv
            \cos^{-1}\p{\uv{s} \cdot \uv{\Omega}_{\rm eff}},
\end{equation}
is an adiabatic invariant.

With this adiabatic invariant, it proves straightforward to understand the final
spin-orbit misalignment angle $\theta$.
First, note that
\begin{equation}
    \cos \theta_{\rm eff}
        = \frac{\uv{s} \cdot \ev{\Omega_{\rm dS}\jhat_{\rm in}}
            - \uv{s} \cdot \jhat_{\rm out} \ev{\Omega_{\rm ZLK}}}{\Omega_{\rm
            eff}},\label{eq:qeff_dot}
\end{equation}
where $\ev{\Omega_{\rm ZLK}} = \Delta \Omega_{\rm in} / P_{\rm ZLK}$, the
average rate of change of $\Omega_{\rm in}$ in an orbital period.
At late times, when $\Omega_{\rm dS} \gg \Omega_{\rm ZLK}$, we see that
$\theta_{\rm eff} \approx \theta$.
Thus, the final spin-orbit misalignment angle can be well-predicted by the
initial value of $\theta_{\rm eff}$ (denoted $\theta_{\rm eff, 0}$) and its
subsequent conservation.

To understand the scalings that govern $\theta_{\rm eff, 0}$,
we will specialize to the case where $\uv{s}_0 \parallel \jhat_{\rm in}$ (a
common assumption, e.g.\ \citealp{LL17, LL18, LL19}).
For the wide binaries studied in the literature, and the ``$90^\circ$
attractor'' \citep{LL18, su2021_lk90}, $\Omega_{\rm dS} \ll \Omega_{\rm ZLK}$
initially, and Eq.~\eqref{eq:qeff_dot} immediately implies that $\cos
\theta_{\rm eff, 0} = \cos I_0$, which is nearly zero for nearly-perpendicular
mutual inclinations of the inner and outer orbits necessary to drive wide
binaries to extreme, merger-capable eccentricities.

\begin{figure}
    \centering
    \includegraphics[width=0.4\columnwidth]{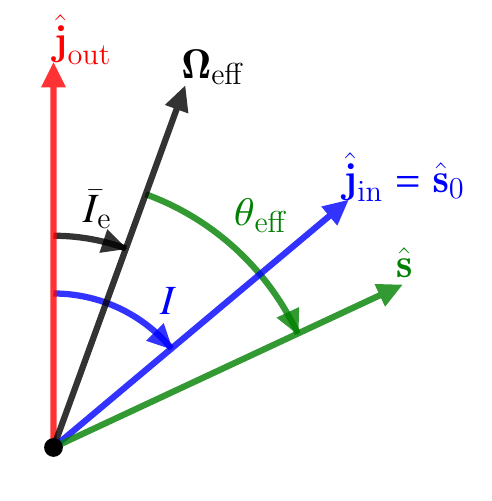}
    \caption{
    Notation of angles used to describe the adiabatic invariant in
    Section~\ref{app:spin_adinvar}.
    $\bm{\Omega}_{\rm eff}$ is the effective spin precession axis, given by
    Eq.~\eqref{eq:dsdt_Weff}.
    Note that initially, $\uv{s}_0 = \jhat_{\rm in}$ , and so $\theta_{\rm eff}
    = \abs{\bar{I}_{\rm e} - I_0}$, as shown.
    }\label{fig:3vec}
\end{figure}
In this paper, we will consider the regime where $\Omega_{\rm dS}$ is not so
much smaller than $\Omega_{\rm ZLK}$ initially, and so we must be more precise.
When also adopting initial spin-orbit alignment (well-justified in our
mechanism, see Section~\ref{s:stellar_evol}), we have that $\uv{s}_0$,
$\ev{\Omega_{\rm dS} \jhat_{\rm in}}$, and $\jhat_{\rm out}$ are all coplanar
(in the co-precessing frame).
Then, since the angle between $\uv{s}_0$ and $\jhat_{\rm out}$ is just $I_0$, it
proves easiest to evaluate the angle between $\ev{\Omega_{\rm dS}
\jhat_{\rm in}}$ and $\jhat_{\rm out}$ ($\bar{I}_{\rm e}$ in the notation of
\citealp{su2021_lk90}) in order to constrain $\theta_{\rm eff, 0}$ (see
Fig.~\ref{fig:3vec}).
This is straightforward:
\begin{align}
    \sin \bar{I}_{\rm e}
        &= \frac{\ev{\Omega_{\rm dS}\hat{\jmath}_{\rm in, \perp}}}{
            \ev{\Omega_{\rm dS}\hat{\jmath}_{\rm in, \parallel}}
            - \ev{\Omega_{\rm ZLK}}},\\
    \theta_{\rm eff, 0}
        &= \abs{\bar{I}_{\rm e} - I_0}.
\end{align}
Here, the $\perp$ and $\parallel$ subscripts denote the components of
$\jhat_{\rm in}$ normal to and parallel to $\jhat_{\rm out}$.
To obtain an even simpler expression at the cost of some accuracy, we can
further approximate $\ev{\Omega_{\rm dS}\jhat_{\rm in}}_{\parallel} \ll
\ev{\Omega_{\rm dS}\jhat_{\rm in}}_{\perp} \ll \Omega_{\rm ZLK}$ (corresponding
to large $I_0$ and weak spin-orbit coupling) and $I_0 \approx 90^\circ$ to
obtain
\begin{align}
    \theta_{\rm eff, 0}
        \simeq{}&
            \abs{\bar{\mathcal{A}}_0 - I_0},\\
    \bar{\mathcal{A}}_0
        \equiv{}&
            \frac{\abs{\ev{\Omega_{\rm dS}\jhat_{\rm in}}}}{
                \ev{\Omega_{\rm ZLK}}}\nonumber\\
        ={}&
            \frac{3Gm_{12}(m_2 + \mu_{\rm in} / 3) a_{\rm out}^3j_{\rm out}^3}{
                2c^2m_3a_{\rm in, 0}^4}\nonumber\\
            &\times \tilde{\mathcal{A}}(a_{\rm in, 0}, I_0,\dots)
            ,\\
    \cos \theta_{\rm eff, 0}
        \simeq{}& \abs{\cos I_0 - \bar{\mathcal{A}}_0}.
\end{align}
where $\tilde{\mathcal{A}}$ is some dimensionless function that primarily
captures the scaling of $\Omega_{\rm dS}$ and $\Omega_{\rm ZLK}$ with $e_{\rm
in, \max}$.
This recovers the key result (Eq.~\ref{eq:naive}) used in the text.

\section{Suppression of ZLK Cycles during the Stellar Phase via Extra Apsidal
Precession}\label{app:SRF}

In this appendix, we delineate the conditions necessary such that a black hole
binary subject to the mechanism proposed in the current work would not have
experienced substantial ZLK-driven coalescence during its stellar phases of
evolution.

In a stellar binary, Eq.~\eqref{eq:milankovich_e} for the evolution of
$\bm{e}_{\rm in}$ has two extra terms due to apsidal precession driven by the
rotational and tidal bulges of the two stars
\citep[e.g.,][]{liu2015suppression}.
Given the combination of the three SRFs (GR, tides, and rotation), the maximum
eccentricity $e_{\rm in, \max}$ that can be attained via the ZLK effect (when $I
= 90^\circ$) is given by
\citep{liu2015suppression}
\begin{align}
        &\frac{\epsilon_{\rm tide, 1}}{15}
            \p{\frac{1 + 3e_{\rm in,\max}^2 + \tfrac{3e_{\rm in,\max}^4}{8}}{j_{\rm in,\min}^9} - 1}
            \nonumber\\
        &{}+
        \frac{\epsilon_{\rm rot, 1}}{3}
            \p{\frac{1}{j_{\rm in,\min}^3} - 1}\nonumber\\
        &{}+
        \epsilon_{\rm GR} \p{\frac{1}{j_{\rm in,\min}} - 1} = \frac{9}{8}e_{\rm in,\max}^2,
        \label{eq:emax_srfs}
\end{align}
where the three dimensionless parameters quantifying the relative strengths of
the general relativistic, tidal, and rotational apsidal precession effects are
given by
\begin{align}
    \epsilon_{\rm GR}
        &= \frac{3Gm_{12}^2 \tilde{a}_{\rm out}^3}{a_{\rm in}^4c^2m_3},\\
    \epsilon_{\rm tide, 1}
        &= \frac{15 m_1m_{12}
            \tilde{a}_{\rm out}^3 k_{2,1}R_1^5}{
                a_{\rm in}^8 m_2m_3},\\
    \epsilon_{\rm rot, 1}
        &=
            \frac{m_{12}\tilde{a}_{\rm out}^3\,
                k_{2,1}R_1^5}{2Ga_{\rm in}^5 m_2m_3}\Omega_{\rm 1s}^2,
\end{align}
and $j_{\rm in, \min} \equiv \sqrt{1 - e_{\rm in, \max}^2}$.
Here, $k_{\rm q, 1} = k_{2, 1}$ is the apsidal motion constant, while $k_{2, 1}$
is the tidal Love number, $\Omega_{1s}$ is the spin frequency of the primary,
and $R_1$ is its radius.
Eq.~\eqref{eq:emax_srfs} can be straightforwardly generalized to include the
contributions from the secondary (i.e.\ $\epsilon_{\rm tide, 2}$ and
$\epsilon_{\rm rot, 2}$).
However, while the combined precessional effects will further suppress
eccentricity excitation, the disparate mass ratios in our considered systems
imply that the SRFs due to either the primary or the secondary (after the
primary has evolved into a BH) will dominate the apsidal precession of the
binary.
As such, it is sufficient to consider only one set of SRFs at a time.

In a stellar binary, the dominant SRF is apsidal precession driven by either the
rotational or tidal bulges of the stars.
For our fiducial parameters, the stellar tidal synchronization timescale due to
the equilibrium tide is \citep{alexander1973weak, hut1981tidal, lai2012}:
\begin{align}
    \frac{1}{t_{\rm s}}
        &= \frac{1}{4k}\frac{3k_2}{Q}
            \p{\frac{m_2}{m_1}}
            \p{\frac{R_1}{a_{\rm in}}}^3
            n_{\rm in}\nonumber\\
        &=
            \frac{1}{1\;\mathrm{Gyr}}
            \p{\frac{2k_2/Q}{10^{-6}}}
            \p{\frac{m_2}{m_1}}
            \p{\frac{R_1}{10R_{\odot}}}^3
            \p{\frac{a_{\rm in}}{2\;\mathrm{au}}}^{-9/2}.
\end{align}
As such, the stars remain rapidly rotating throughout their MS lifetimes
($\lesssim \mathrm{Myr}$), and the rotational bulge provides the dominant source
of apsidal precession to truncate the ZLK cycles.
Note that, when other SRFs are negligible, Eq.~\eqref{eq:emax_srfs} reduces to
\begin{equation}
    \frac{8\epsilon_{\rm rot}}{27}
        = \frac{j_{1,\min}^3\p{1 + j_{1,\min}}}{
            1 + j_{1,\min} + j_{1,\min}^2}.\label{eq:jmin_rot}
\end{equation}
Thus, we conclude that $\epsilon_{\rm rot} \geq 9/4$ suppresses ZLK oscillations
entirely ($j_{1, \min} = 1$; see Fig.~6 of \citealp{liu2015suppression}). For
our fiducial parameters,
\begin{align}
    \epsilon_{\rm rot, 1}
        ={}& 2.3 \frac{m_{1,\star}}{m_{2,\star}}
            \p{\frac{k_{q, 1}}{0.02}}
            \p{\frac{m_{12,\star}}{100M_\odot}}
            \p{\frac{\tilde{a}_{\rm out}}{6400\;\mathrm{au}}}^3
                \nonumber\\
        &\times \p{\frac{R_1}{10 R_{\odot}}}^2
            \p{\frac{a_{\rm in, \star}}{2\mathrm{au}}}^{-5}
            \p{\frac{m_3}{10^7M_{\odot}}}^{-1}\nonumber\\
        &\times \p{\frac{\Omega_{\rm s, 1} / \Omega_{\rm dyn, 1}}{0.8}}^2,
\end{align}
where $\Omega_{\rm dyn, 1} \equiv \sqrt{Gm_{1,\star}/R_1^3}$ is the dynamical frequency
of the primary and is also its maximal spin frequency.
We have taken values for the stellar structure from numerical simulations using
Modules for Experiments in Stellar Astrophysics \citep[MESA][]{MESA_1, MESA_2,
MESA_3, MESA_4, MESA_5, MESA_6} representative of a $60M_{\odot}$
star for solar and sub-solar metallicities along its main sequence.
As such, we conclude that the rotational bulge of the primary is able to
suppress ZLK oscillations for our fiducial parameters.
As will become clear below, this suppression is active until both stars collapse
to BHs.

To complete the picture, once the stellar binary forms a BBH, the only SRF is
the apsidal precession due to first-order post-Newtonian effects.
In the absence of other SRFs, Eq.~\eqref{eq:emax_srfs} reduces to
\begin{equation}
    \frac{8\epsilon_{\rm GR}}{9}
        = j_{1,\min}\p{1 + j_{1,\min}}.\label{eq:jmin_gr}
\end{equation}
Thus, we find that $\epsilon_{\rm GR} \gtrsim 9/4$ suppresses ZLK entirely
(see again Fig.~6 of \citealp{liu2015suppression}).
For fiducial parameters, $\epsilon_{\rm GR}$ evaluates to
\begin{align}
    \epsilon_{\rm GR}
        ={}& 0.12
            \p{\frac{m_{\rm 12}}{50M_\odot}}^2
            \p{\frac{\tilde{a}_{\rm out}}{6400\;\mathrm{au}}}^3\nonumber\\
        &\times \p{\frac{a_{\rm in}}{2\mathrm{au}}}^{-4}
            \p{\frac{m_3}{10^7M_{\odot}}}^{-1}.
\end{align}
Note that $m_{12} < m_{12, \star}$ due to mass loss during binary stellar
evolution.
Thus, we find that eccentricity excitation via the ZLK effect can be achieved
during the BH phase even as it is suppressed during the stellar phase.

\begin{adjustwidth}{-\extralength}{0cm}

\reftitle{References}


\bibliography{refs}

%


\PublishersNote{}
\end{adjustwidth}
\end{document}